\documentclass[ps,twocolumn,superscriptaddress,citeautoscript,floatfix,showpacs,reprint]{revtex4-1}
 
\pdfoutput=1
\usepackage{natbib}
\usepackage{microtype}
\usepackage[english]{babel}
\usepackage{amsmath}
\usepackage{amssymb}
\usepackage{amsfonts}
\usepackage{graphicx}
\usepackage{bm}
\usepackage{complexity}
\usepackage{framed}
\usepackage{enumitem}

\graphicspath{{images/}}

\usepackage[dvipsnames]{xcolor}
\usepackage{xcolor,colortbl}
\definecolor{darkblue}{rgb}{0.1,0.2,0.6}
\definecolor{darkred}{rgb}{0.8,0.1,0.2}
\definecolor{Gray}{gray}{0.9}
\definecolor{Black}{gray}{1.0}

\usepackage[colorlinks,citecolor=darkblue,linkcolor=darkred,urlcolor=darkblue]{hyperref}
\usepackage[all]{hypcap} 
\usepackage{cleveref}

\newcommand{\ket}[1]{\ensuremath{\left|#1\right\rangle}} 
\newcommand{\bra}[1]{\ensuremath{\left\langle#1\right|}} 
\DeclareMathOperator{\Avg}{Avg} 
\DeclareMathOperator{\Stdev}{StdDev} 

\newcommand{\ie}{{\it i.e.\ }}   
\newcommand{\eg}{{\it e.g.\ }}   

\def\dist{\operatorname{dist}}

\def\rem{\mathrm{rem}}
\def\sat{\mathrm{sat}}

\def\qu{\mathrm{qu}}
\def\cl{\mathrm{cl}}

\def\fin{\mathrm{fin}}
\def\dyn{\mathrm{dyn}}
\def\st{\mathrm{st}}

\newcommand{\iprod}[2]{\langle #1 \vert #2 \rangle}

\graphicspath{{./figures/}}

\begin{document}

\title{Multifractal Dynamics of the QREM}

\def\quail{
	Quantum Artificial Intelligence Lab. (QuAIL), 
    NASA Ames Research Center, Moffett Field, CA 94035, USA
}

\def\kbr{
	KBR, 
    601 Jefferson St., Houston, TX 77002, USA
}

\def\sissa{
	International School for Advanced Studies,
    Via Bonomea 265, 34136 Trieste TS, Italy
}

\def\ictp{
	International Centre for Theoretical Physics,
   Strada Costiera 11, 34151 Grignano TS, Italy
}

\def\usra{
	Universities Space Research Association,
    7178 Columbia Gateway Drive, Columbia, MD 21046, USA
}

\def\infn{
	Istituto Nazionale di Fisica Nucleare,  Sezione  di  Trieste,
	34136  Trieste TS,  Italy
}
    
\author{Tommaso Parolini}
	\email{tparolin@sissa.it}
	\affiliation{\quail}
	\affiliation{\usra}
	\affiliation{\sissa}
	\affiliation{\ictp}
	\affiliation{\infn}

\author{Gianni Mossi}
	\email{gianni.mossi@nasa.gov}
	\affiliation{\quail}
	\affiliation{\kbr}

\date{\today}

\begin{abstract}
We study numerically the population transfer protocol on the Quantum Random Energy Model and its relation to quantum computing, for system sizes of $n\leq 20$ quantum spins. We focus on the energy matching problem, \ie finding multiple approximate solutions to a combinatorial optimization problem when a known approximate solution is provided as part of the input. We study the delocalization process induced by the population transfer protocol by observing the saturation of the Shannon entropy of the time-evolved wavefunction as a measure of its spread over the system. The scaling of the value of this entropy at saturation with the volume of the system identifies the three known dynamical phases of the model. In the non-ergodic extended phase, we observe that the time necessary for the population transfer to complete follows a long-tailed distribution. We devise two statistics to quantify how effectively and uniformly the protocol populates the target energy shell. We find that population transfer is most effective if the transverse-field parameter $\Gamma$ is chosen close to the critical point of the Anderson transition of the model. In order to assess the use of population transfer as a quantum algorithm we perform a comparison with random search. We detect a ``black box'' advantage in favour of PT, but when the running times of population transfer and random search are taken into consideration we do not see strong indications of a speedup at the system sizes that are accessible to our numerical methods. We discuss these results and the impact of population transfer on NISQ devices.
\end{abstract}

\maketitle

\section{Introduction}\label{sec:introduction}

Quantum tunnelling has been for a long time claimed to be the mechanism behind a possible quantum speedup in quantum optimization algorithms. This was a particularly prominent claim in the early days of Adiabatic Quantum Computing (AQC). More recently, a new approach --- named Population Transfer (PT) by its authors \cite{Smelyanskiy2020}  --- was proposed that uses coherent dynamics of quantum disordered systems in order to explore the rough energy landscape of combinatorial optimization problems in a better way than what is obtainable through classical local search algorithms. PT does not aim at solving combinatorial optimization problems, \ie finding good approximate solutions where none are known. Instead, PT is designed to take an approximate solution (a \emph{seed}) that the user is expected to provide, and use it to find other approximate solutions whose energy is comparable to the energy of the original one.

In order to use PT, one starts with a classical (\ie $\sigma^z$-diagonal) energy term $H_p$ and modifies it by adding of a transverse-field term in order to introduce a non-trivial quantum dynamics.
\begin{equation} \label{eq:hamiltonian}
H_\Gamma = H_p - \Gamma \sum_{i=1}^n \sigma_i^x
\end{equation}
In the standard setup, the $H_p$ term encodes the optimization problem that one is interested in studying. The PT protocol start with a given binary string $z_0$ and proceeds as follows.
\begin{enumerate}
	\item Prepare the state $\ket{z_0}$.
	\item Fix a value $\Gamma$ and a final time $t_{\fin}$, and evolve the system according to the propagator generated by the Hamiltonian of Eq.~\eqref{eq:hamiltonian}, obtaining the state $\ket{\psi(t_{\fin})} = e^{-iH_\Gamma t_{\fin}}\ket{z_0}$.
	\item Perform a measurement in the computational basis and obtain a classical state $z$ with probability
	$$p(z) = \lvert \iprod{z}{\psi(t_{\fin})} \lvert^2.$$
\end{enumerate}
In order to provide PT with a well-defined computational objective, one can define an ``energy matching'' problem~\cite{Baldwin2018}. Fix a target energy density $\epsilon = E/n$ and let $g(n)$ a positive function of infinitesimal order $o(n^0)$; then we define

\begin{framed}
	\begin{itemize}[leftmargin=0pt]
		\item[]{\textsc{Problem class}: \( g(n) \)-energy matching for \( H_p \)}
		\item[]\textbf{Input:} an instance \( H_p \) over \( n \) binary variables and an \( n \)-bit binary string \( z_0 \) with \( H_p(z_0) = E \).
		\item[]\textbf{Output:} an \( n \)-bit binary string \( z \neq z_0 \) such that \( E(1-g(n)) \leq H_p(z) \leq E(1+g(n)) \).
	\end{itemize}
\end{framed}
According to this definition, $E$ is the target energy of the PT protocol and $g(n)$ defines the approximation one is willing to accept in the solutions. If one defines an ``energy shell'' around $E$ of width $2 \Delta E$
$$
\Omega(E,\Delta E) \equiv \big\{ z \bigm\vert E-\Delta E \leq H_p(z) \leq E+\Delta E \big\}
$$
then the \emph{target states} for the energy matching problem are the states in the ``punctured'' energy shell, $z \in \Omega_n \equiv \Omega(E,g(n)) \setminus \{z_0\}$.
In the following we will always use the function $g(n) = (10n)^{-1/2} $, so that at \( n = 10 \) the shell's width relative to the target energy is 10\%, narrowing down as $n\rightarrow \infty$ to pinpoint the energy density $\epsilon$. We will therefore drop the $g(n)$- prefix and just use ``energy matching'' throughout.

The efficacy of PT on solving the energy matching problem depends on the choice of the two user-defined parameters $\Gamma$ and $t_{\fin}$ that characterize the PT dynamics. One of the aims of this paper will be to study how this choice can be made properly. In the following we study numerically the PT protocol in the Random Energy Model (REM) for up to $n = 20$ quantum spins. In Section \ref{sec:literature} we briefly review the REM and some results relative to its quantum version (QREM) that are relevant to our work. In Section \ref{sec:dynamics} we focus on a low-energy shell and use the Shannon entropy of the energy eigenstates of the QREM Hamiltonian (as a measure of the size of their support) to detect the Anderson and the Ergodic transitions as $\Gamma$ is increased from zero. Having identified the non-ergodic extended (NEE) regime, we study the dynamics of the relaxation process induced by PT as the initial state hybridizes with other computational-basis states. This relaxation time is the running time of PT as a quantum algorithm for energy matching, and we find it to be exponentially increasing with the system size. In Section \ref{sec:algorithm} we analyse the performance of the PT protocol as a solution-mining quantum algorithm. We observe that if one measures a PT-evolved wavefunction in the computational basis, then the probability of finding a target state is asymptotically better than what is achieved by random search. However, if the time required to produce such a state is taken into account, then PT and random search scale approximately in the same way at the system sizes we can access numerically, even though we are unable extrapolate this result to the asymptotic regime due to strong finite-size effects. In Section \ref{sec:conclusions} we summarize our findings and propose further lines of research.

\section{Classical and Quantum REM}\label{sec:literature}

We define the Random Energy Model by the $\sigma^z$-diagonal Hamiltonian
\begin{equation}\label{eq:rem}
H_{\rem} = \sum_{z=0}^{2^n-1} E_z \ket{z}\bra{z}
\end{equation}
where the $2^n$ energies $E_z$ are i.i.d.\ Gaussian random variables distributed according to $p(E) = \sqrt{1/\pi n} \exp(-E^2/n)$. The classical Random Energy Model was introduced in Ref.~\cite{Derrida1981} by Derrida in order to provide a simplified model that exhibits some of the glassy physics of the Sherrington--Kirkpatrick spin glass \cite{Sherrington1975}. Being a collection of states with uncorrelated energies, the number of states at a given energy is given in the thermodynamic limit by
$$
\rho(E) = \frac{2^n}{\sqrt{n\pi}}  e^{-E^2/n},
$$
The REM is the prototypical example of a model with ``golf-course'' energy landscape, where the typical energy density is $\epsilon \equiv E/n =0$ (the flat ``plateau'') while states with finite energy densities $0 < \lvert \epsilon \rvert \leq \sqrt{\log 2}$ are randomly scattered inside of the plateau, and extensively separated between one another (with Hamming distance $d = O(n)$). In this sense, the classical random energy model exhibits energy clustering \cite{Mezard2009, Bapst2013}, albeit of a degenerate kind as each ``cluster'' is composed of a single state. In the large-$n$ and large-time limit, its local dynamics is exactly captured by Bouchaud's trap model \cite{Bouchaud1992,Gayrard2016,Baity-Jesi2018}.

The Quantum Random Energy Model (QREM) is the transverse-field model obtained by taking the REM Hamiltonian as the problem Hamiltonian ($H_{\rem} = H_p$ in Eq.~\eqref{eq:hamiltonian}).
The QREM Hamiltonian is formally equivalent to the Anderson model with a Gaussian-distributed disordered local potential and nearest-neighbour hopping defined over the Boolean hypercube of connectivity $n$, where the $\sigma^z$-diagonal term $H_{\rem}$ plays the role of the random local potential and the transverse field is the kinetic term. The number of sites on this fictitious lattice (that we call the \emph{volume} of the system and indicate with $V$) is equal to $2^n$. Unlike the standard Anderson model, where the hopping coefficient is kept fixed and the transition is induced by the disorder-strength parameter $W$, in our approach it is more common to keep the energy scale of the local potential fixed, and transitions are induced by controlling the kinetic energy parameter $\Gamma \sim 1/W$. This allows one to study the dynamics of the PT protocol through the methods of Anderson localization.

\begin{figure}
  \centering
    \label{fig:qrem_phase_diagram}
    \includegraphics[width=0.45\textwidth]{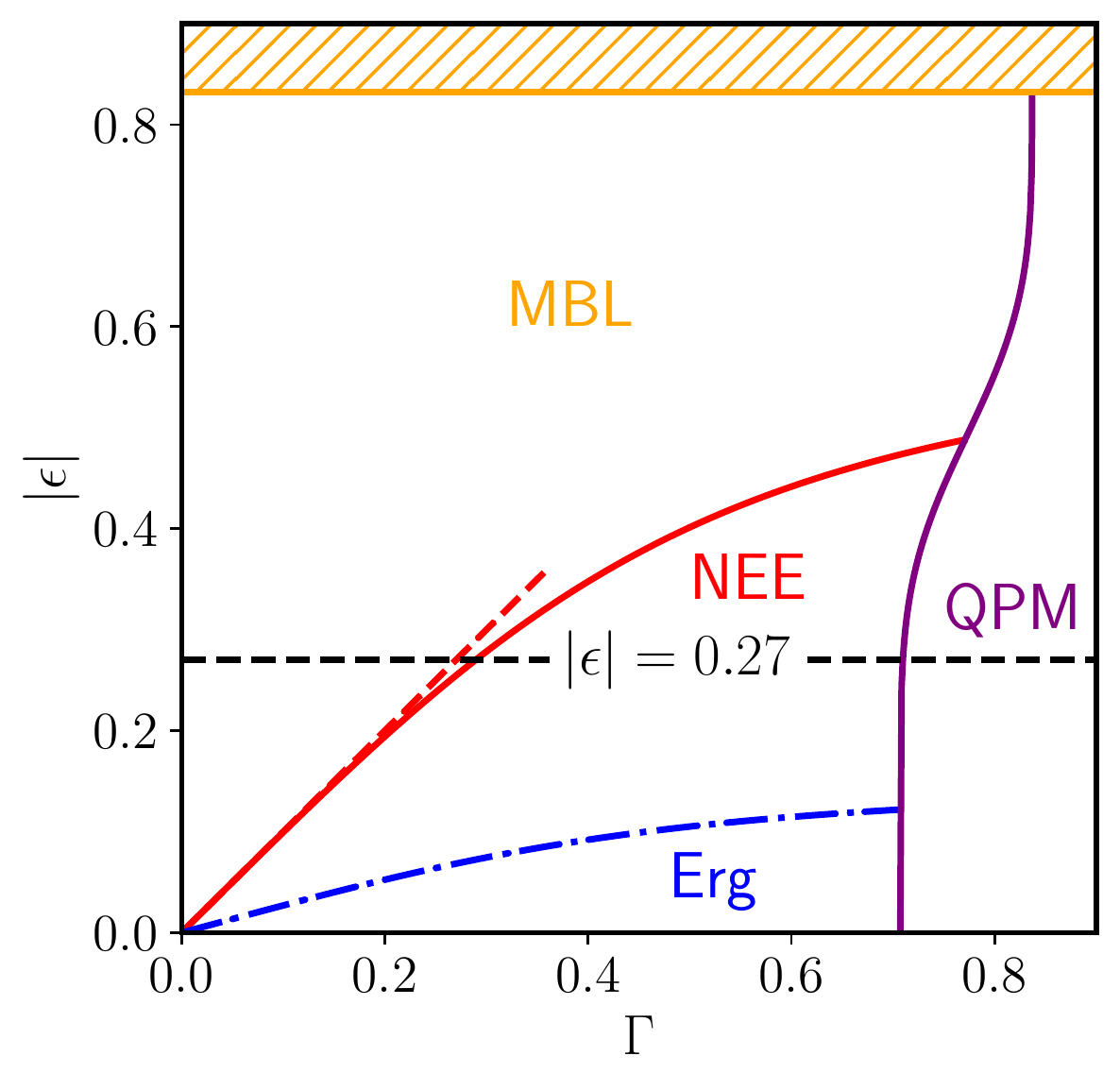}
    \caption{Dynamical phase diagram of the QREM. The solid red line \( \Gamma_\textrm{MBL}(\epsilon) \) marks the Anderson transition; the dashed red line represents the leading-order estimate \( \Gamma_\textrm{MBL}(\epsilon) \approx |\epsilon| \) in the region where the approximation is at least 90\% accurate. The orange line is the band edge. The purple line denotes the onset of the quantum paramagnetic region. The dot-dashed blue line represents a rough estimate of the putative nonergodic--ergodic transition. Our analysis is carried out along the dashed black line, \( |\epsilon| = 0.27 \).}
\end{figure}

The dynamical phase diagram of the QREM has been the object of study of several previous works. In Ref.~\cite{Baldwin2016,PhysRevLett.113.200405} the mobility edge for the Anderson transition was estimated as $\Gamma_c(\epsilon) = \epsilon$ using the forward scattering approximation. This is generally believed to be correct near $\Gamma = 0$ but possibly unreliable at larger $\Gamma$ as it is involves a perturbative calculation of the Green's function with $\Gamma$ as a small parameter. In Ref.~\cite{Faoro2019} it was argued that at low enough energies and at least in an interval of $\Gamma$, the Rosenzweig--Porter (RP) model should provide an effective model of the QREM. Through this mapping one finds that the QREM should exhibit all of the three dynamical phases of the RP model, \ie localized, extended nonergodic and extended ergodic. The two transitions between these phases are respectively called the Anderson and the Ergodic transitions. In Ref.~\cite{Smelyanskiy2019} the authors derive an estimate of the Ergodic transition that coincides with the one in Ref.~\cite{Faoro2019}, but argue that the NEE phase is layered in an alternating sequence of two distinct subphases. The different estimates for the three phase transitions for the QREM are summarized in Fig.~\ref{fig:qrem_phase_diagram}.

As for the population transfer protocol, following the analysis of \cite{Smelyanskiy2020} on the impurity band model (an analytically tractable  model related to the QREM) one would expect a polynomial speedup over random search, although a more recent work by the same authors \cite{Smelyanskiy2019} provides a precautionary statement against naively drawing that conclusion. In a different work on the quantum $p$-spin model \cite{Baldwin2018}, the PT dynamics is computed through the Schrieffer--Wolff perturbation theory in $\Gamma$. Their analysis (extrapolated to the QREM by taking the standard limit $p\rightarrow\infty$) claims that no speedup is expected on the QREM.

\section{Quench Dynamics in the nonergodic Phase}\label{sec:dynamics}

In this section we study the dynamics of the PT protocol described above in the following steps. 1) We detect the transition from a localized to an extended phase using the volume scaling of the Shannon entropy of the energy eigenfunctions in the computational basis. We observe non-ergodicity for a significant interval of $\Gamma$ values. 2) We define a criterion for estimating the timescale $t_\sat$ of the PT-induced delocalization process by observing the saturation of the Shannon entropy of the time-evolved wavefunction. We study the volume scaling of the saturated entropy and compare it with the scaling of entropy of the energy eigenfunctions. 3) We study the distribution of the saturation times $t_\sat$ over the disorder.

\subsection*{Entropy of the energy eigenstates}
\begin{figure}
  \centering
    \label{fig:D_scaling_exponent}
    \includegraphics[width=0.5\textwidth]{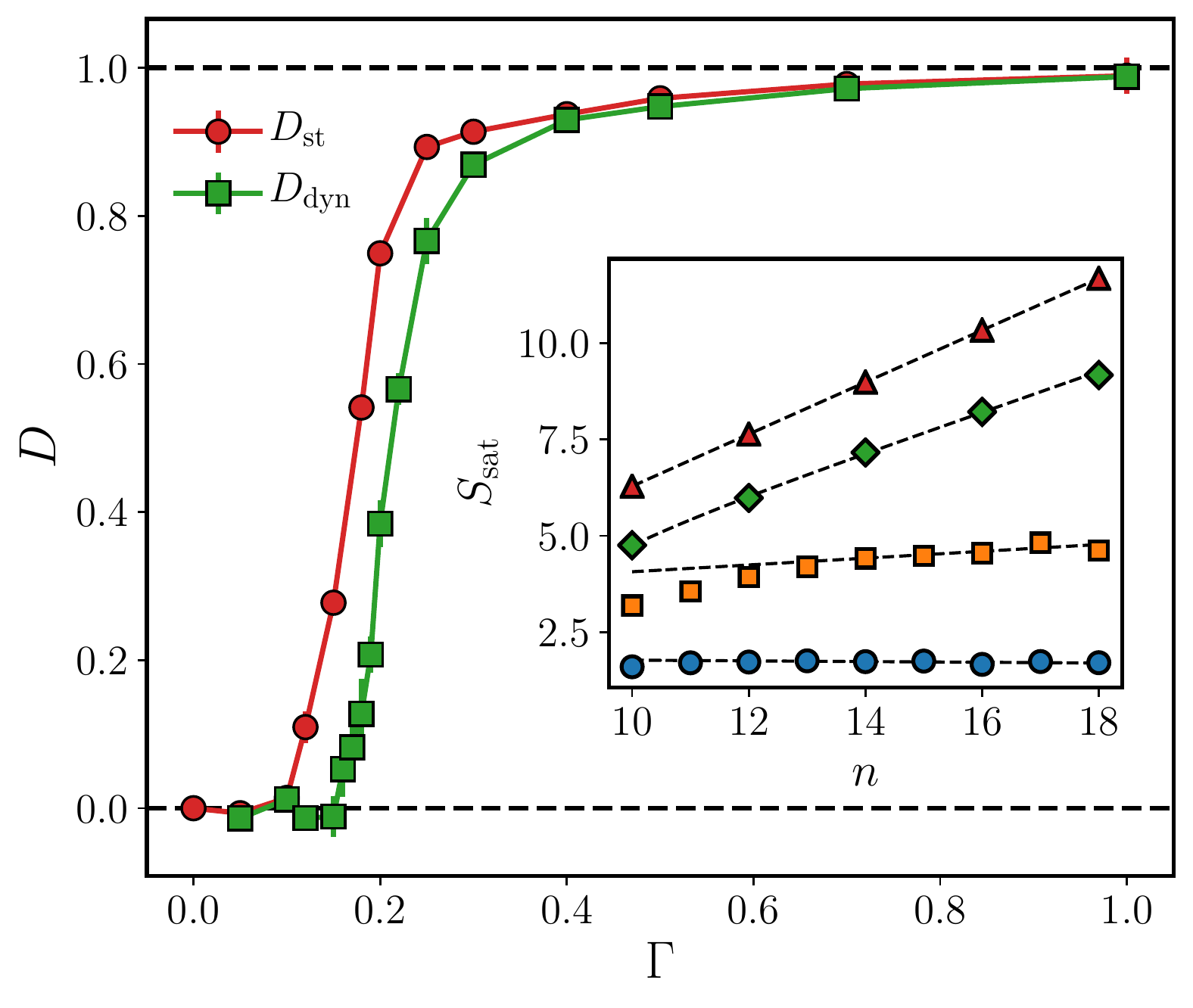}
    \caption{Behavior of the scaling dimension of the eigenstate entropy \( D_\mathrm{st} \) (red circles) and of the dynamical entropy \( D_\mathrm{dyn} \) (green squares). \emph{Inset:} linear fit of the dynamical entropy for fixed values of the transverse field (from bottom to top, \( \Gamma = 0.05, 0.16, 0.2, 0.4 \)).}
\end{figure}

Multiple ways of detecting the NEE phase have been proposed in the literature (see \eg Refs.~\cite{Kravtsov2015,Facoetti2016,Altshuler2016nonergodic,Pino2017,Kravtsov2018,Faoro2019}). We found that the easiest way of detecting the Anderson and the Ergodic transitions numerically is the scaling analysis of the Shannon entropy of the energy eigenstates $\ket{\psi_\alpha}$ of the QREM Hamiltonian in Eq.~\eqref{eq:hamiltonian}. Given any such eigenstate, its Shannon entropy (in the computational basis $\{z\}$) is
\begin{equation}\label{eq:energy_eigenfunction_entropy}
S[\psi_\alpha] \equiv - \sum_z \lvert \iprod{z}{\psi_\alpha} \rvert^2 \log \lvert \iprod{z}{\psi_\alpha} \rvert^2,
\end{equation}
and one defines the scaling dimension \footnote{note that in the multifractal literature (see \eg Ref.~\cite{RevModPhys.80.1355}), this scaling dimension $D_{\st}$ coincides with the fractal dimension $D_1$, also known as the ``information dimension''. However, we will make no use of this connection in the present work.} $D_{\st}$ through its asymptotic behaviour
\begin{equation}\label{eq:d_st}
S[\psi_\alpha] \sim D_{\st} \log(V).
\end{equation}
Since $0 \leq S \leq \log V$, the exponent $D_{\st}$ must lie between zero and one. Note that $\exp(S)$ defines the size of the ``typical set'' that includes the sites $z$ that the wavefunction associates with higher probabilities $\lvert \psi_z \rvert^2$, also called the ``support set'' of the wavefunction in the physics literature \cite{1401.0019,PhysRevLett.113.046806}. An exponent $D_{\st}$ indicates that the number of states in the support set of the wavefunctions scales as $V^{D_{\st}}$ in the large $V$ limit. In the localized regime the energy eigenfunctions decay exponentially away from their localization center so that most of the amplitude is concentrated in a region of size $O(1)$, while in the ergodic extended regime they are roughly uniformly extended over the whole system (of size $O(V))$. This means that the possible values of $D_{\st}$ can be used to identify the dynamical phases in the following way:
$$
\begin{cases} D_{\st}=0 & \text{localized phase}\\
0<D_{\st}<1 & \text{nonergodic extended phase}\\
D_{\st}=1 & \text{ergodic extended phase.}
\end{cases}
$$
Thus, in order to assign a point $(\epsilon,\Gamma)$ in the phase diagram to one of the three dynamical phases, we do as follows. For each system size $n = 8,\ldots,18$ we generate a large number of disorder realizations $H_{J}$ of the classical REM model of Eq.~\eqref{eq:rem}. Then, we create the QREM realization $H_J(\Gamma)$ by adding a transverse-field term with the given value of $\Gamma$ and we use the shift-invert method to extract the energy eigenstate $\ket{\psi_{\alpha}}$ of the QREM Hamiltonian whose energy is closest to $E = \epsilon n$. We compute the Shannon entropy of the eigenstate $\ket{\psi_{\alpha}}$ in the $\sigma^z$ basis (Eq.~\eqref{eq:energy_eigenfunction_entropy}) and take the average of the disorder. Thus, we obtain the typical values $S_V$ for systems of volumes $V= 2^8,2^9,\ldots,2^{18}$. Finally, through Eq.~\eqref{eq:d_st} one can see that a linear fit of $S_V$ vs $\log V$ (for large enough $V$) will produce the $D_{\st}$ associated to the phase-diagram point $(\epsilon,\Gamma)$.
The results for an energy density $\epsilon = 0.27$ and $0 \leq \Gamma \leq 1$ are shown in Fig.~\ref{fig:D_scaling_exponent}, where one notes that $D_{\st}$ (indicated by the red circles) increases continuously from zero to one as $\Gamma$ is increased from zero. 

\subsection*{Dynamical Entropy}

Since Anderson's seminal paper \cite{Anderson1958}, the Anderson transition was defined by two qualitatively different dynamical behaviours that can occur in a quantum system as an initially localized state is left to evolve under coherent evolution. 

In this section we study numerically this delocalization process in the QREM, with an emphasis on the NEE phase. In analogy with what we did in the previous section, we will be focussing on the dynamical value of the Shannon entropy of the wavefunction $\ket{\psi(t)} = \exp(-iHt)\ket{z_0}$, that is 
\begin{equation}\label{eq:wavefunction_entropy}
S(t) \equiv - \sum_z \lvert \iprod{z}{\psi(t)} \rvert^2 \log \lvert \iprod{z}{\psi(t)} \rvert^2.
\end{equation}

At time $t=0$ the wavefunction is fully concentrated on the initial state and its entropy is therefore $S=0$. The entropy then increases as the quantum dynamics populates resonant states, up to an (eventual) stable value $S_\sat$.

For our purposes, an instance is considered to have ``saturated'' once the distribution of the instantaneous values of the entropy becomes in a sense indistinguishable from Gaussian noise around a stable mean. More precisely, suppose that the entropy is being sampled at times \( t \in \{t_1 < t_2 < \dotsb < t_\mathrm{M}\} \). Now call \( S[t_j:t_M] = \left(S(t_j), S(t_{j+1}), \dotsc, S(t_M)\right) \) the sequence of entropy snapshots after time \( t_j \) for a certain instance. We say that the instance ``saturates at time \( t_J \)'' if \( t_J \) is the smallest time such that the values \( S(t_{J-1}), S(t_{J-2}), \dotsc, S(t_{J-k}) \) are \emph{all} less likely than a given probability threshold to have been sampled from a normal distribution with compatible average and variance (the arbitrary parameter \( k > 1 \) is used to rule out genuinely random large deviations from the mean).

In other words, for a given small \( p > 0 \) we compute a corresponding threshold \( \xi = \sqrt{2}\mathrm{Erf}^{-1}(1-p) \), such that a standard Gaussian variable has probability \( p \) of being larger than \( \xi \) in absolute value. Then, we look for the smallest \( J \) such that
\begin{equation}\label{eq:saturation-condition}
    \phi(t_j) \equiv \frac{\left|S(t_j)-\Avg{S[t_J:t_M]}\right|}{\Stdev{S[t_J:t_M]}} > \xi
\end{equation}
for all \( j \in \{J-k, \dotsc, J-1\} \), where \( \Avg{X} \) represents the arithmetic mean of process \( X = (x_1, \dotsc, x_K) \) and \( \Stdev{X} = \sqrt{\frac{K}{K-1}\Avg{\left[\left(X-\Avg{X}\right)^2\right]}} \). We take \( k = 4 \) (3 for \( n = 10, 11 \)), \( p = 1\% \) and require that the number of samples between \( t_J \) and \( t_M \) be sufficiently large (at least 15 samples).

The above condition allows us in principle to define \( t_J \) as the saturation time. However, due to the finite sampling rate, this will result in an ``aliased'' distribution for the saturation time, with the loss of precision associated to constraining \( t_J \) to belong to the set of sampled times \( {\{t_j\}}_{j=1}^M \). In order to ameliorate this effect we introduce a simple interpolation procedure.

Suppose that \( t_J \) is the smallest sampled time where the aforementioned condition holds. Then by assumption \( \phi(t_J) \le \xi < \phi(t_{J-1}) \). Assuming that function \( \phi \) is continuous, we may approximate its unsampled behavior between \( t_J \) and \( t_{J-1} \) through a linear function \footnote{We actually use a linear-in-\( \log(t) \) function as our sampled times are logarithmically spaced.}, and define the saturation time \( t_\mathrm{sat} \) through the condition
\begin{equation*}
    \phi(t_\mathrm{sat}) = \xi,
\end{equation*}
with the solution obviously satisfying \( t_{J-1} < t_\mathrm{sat} \le t_J \). This will give us a more natural distribution for the saturation time, as \( t_\mathrm{sat} \) is now unconstrained.

\begin{figure}
  \centering
    \label{fig:saturation_condition}
    \includegraphics[width=0.5\textwidth]{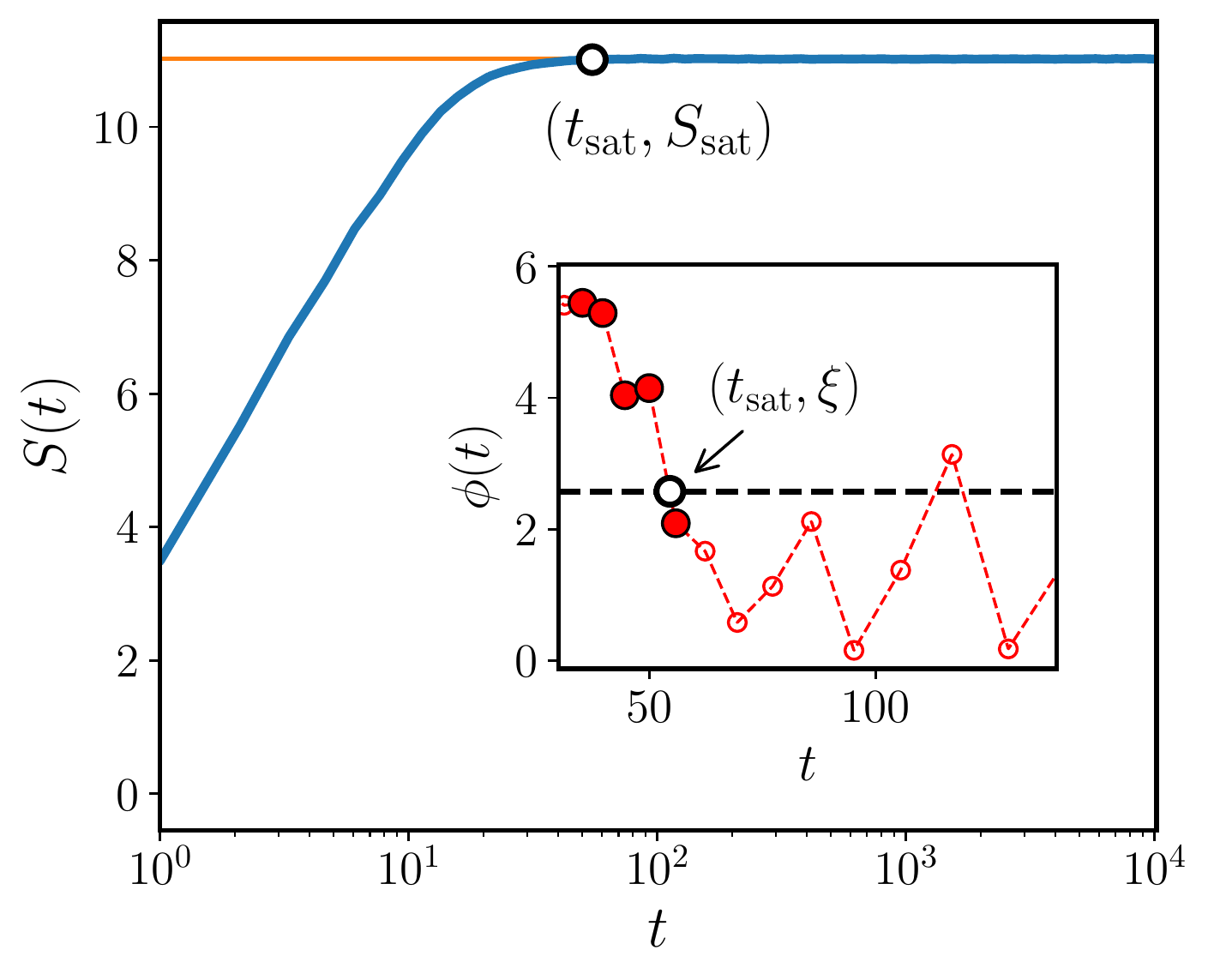}
    \caption{Typical behaviour of the Shannon entropy $S(t)$ of the wavefunction (Eq.~\eqref{eq:wavefunction_entropy}) in the PT dynamics for a single instance with \( n = 18 \). $S$ starts at zero and increases with $t$, up to a point of saturation $S_{\sat} = S(t_\mathrm{sat})$ (orange line), where it settles with small (indiscernible in the plot) random fluctuations, which we model as a Gaussian process. \emph{Inset:} instantaneous value of the discrepancy \( \phi(t) \), see Eq.~\eqref{eq:saturation-condition}, in a limited range of times. Highlighted in red are the points \( t_{J-4}, \dotsc, t_J \), which define the saturation condition as explained in the text. Note the genuinely random large deviation at \( t \approx 120 \), which is correctly ignored by having \( k > 1 \).}
\end{figure}

Fig.~\ref{fig:saturation_condition} portrays the typical time dependence of \( S(t) \) and provides a visualization of the definition of \( t_\sat \) in terms of the function \( \phi(t) \).

\subsection*{$S_{\sat}$: fractal delocalization of the wavefunction}
In analogy with the previous Section, we use the saturation values of the Shannon entropy $S_{\sat}$ to measure the size of the region of the lattice that the wavefunction delocalizes over under unitary evolution. In the ergodic phase one expects that for large $t$, $S(t) \sim \log(V)$ as the wavefunction eventually populates all available volume, while in a strongly localized phase (\eg the localized phase of the Anderson model in finite dimensions, where the RAGE theorem holds \cite{aizenman2015random}), $S(t) \sim O(1)$ since the wavefunction is trapped for all times in a finite region of space surrounding the initial position. According to \cite{Smelyanskiy2020}, in the NEE phase the initial state should delocalize over the common support (\ie the intersection of the roughly-similar supports) of the energy eigenstates belonging to an energy miniband, states that we previously saw have fractal supports. The intersection of fractal sets is commonly a fractal set itself \footnote{see \eg Ref \cite{falconer2013fractal}, chapter 8 for details.} even though it may be characterized by a smaller scaling dimension than the one obtained from the original sets (this is to be expected: the intersection of a collection of sets $A = \bigcap_{\alpha} A_\alpha$ is generically smaller than any set $A_\alpha$).

We define a scaling dimension $D_{\dyn}$ for the saturation values of the Shannon entropy of the wavefunction, in a way analogous to the static $D_{\st}$ of the previous section:
\begin{equation}
    S_V(t_\infty) \sim D_\mathrm{dyn} \log V
\end{equation}
where \( S_V \) is the median entropy of the time-evolved wavefunction for a system of volume \( V \) and we sample at a large time \( t_\infty \) such that \( t_\infty \gg t_\mathrm{sat} \) for every instance.

For large enough values of the transverse field, the median saturated entropy \( S_\mathrm{sat} \) is nicely fit by a law of the form \( S_V(t_\infty) = D_\mathrm{dyn}\log{V} + C_0 + C_{-1}/V \) (cf.~\cite{Kravtsov2015}), where the finite-size deviation coefficient \( C_{-1} \) is always comfortably small.

However, as can be appreciated from the inset of Fig.~\ref{fig:D_scaling_exponent} (\( \Gamma = 0.16 \), square orange markers), the small-\( \Gamma \) data are affected by larger finite-size effects in that they take longer to reach the asymptotic regime. These effects are not well captured by a continuous ansatz, but are better described as a sudden change of regime. Combined with the small value of the scaling dimension in that regime, this requires special care in extracting a useful estimate of \( D_\mathrm{dyn} \). We chose to restrict our fitting procedure to the \( n \ge 14 \) data for \( 0.12 \le \Gamma \le 0.20 \), whereas for \( \Gamma < 0.12 \) the data is essentially too flat to detect a regime change and all sizes can be fitted. In both cases, we use a function \( D_\mathrm{dyn}\log{V} + C_0 \) for the fit.

With these provisions, \( D_\mathrm{dyn} \) can be computed and compared to \( D_\mathrm{st} \), as shown in Fig.~\ref{fig:D_scaling_exponent}. The dynamical exponent has a behaviour qualitatively identical to the static one, albeit shifted in the \( \Gamma \) axis, and it can be seen to satisfy
\begin{equation*}
    D_\mathrm{dyn} \le D_\mathrm{st}.
\end{equation*}
This stands in agreement with our above intuition that the support of the time-evolved wavefunction essentially consists of the intersection of multiple eigenstate supports, themselves scaling in size with fractal exponent \( D_\mathrm{st} \).

\subsection*{Decay times for delocalization} 
\begin{figure}
  \centering
    \label{fig:teq_histograms}
    \includegraphics[width=0.5\textwidth]{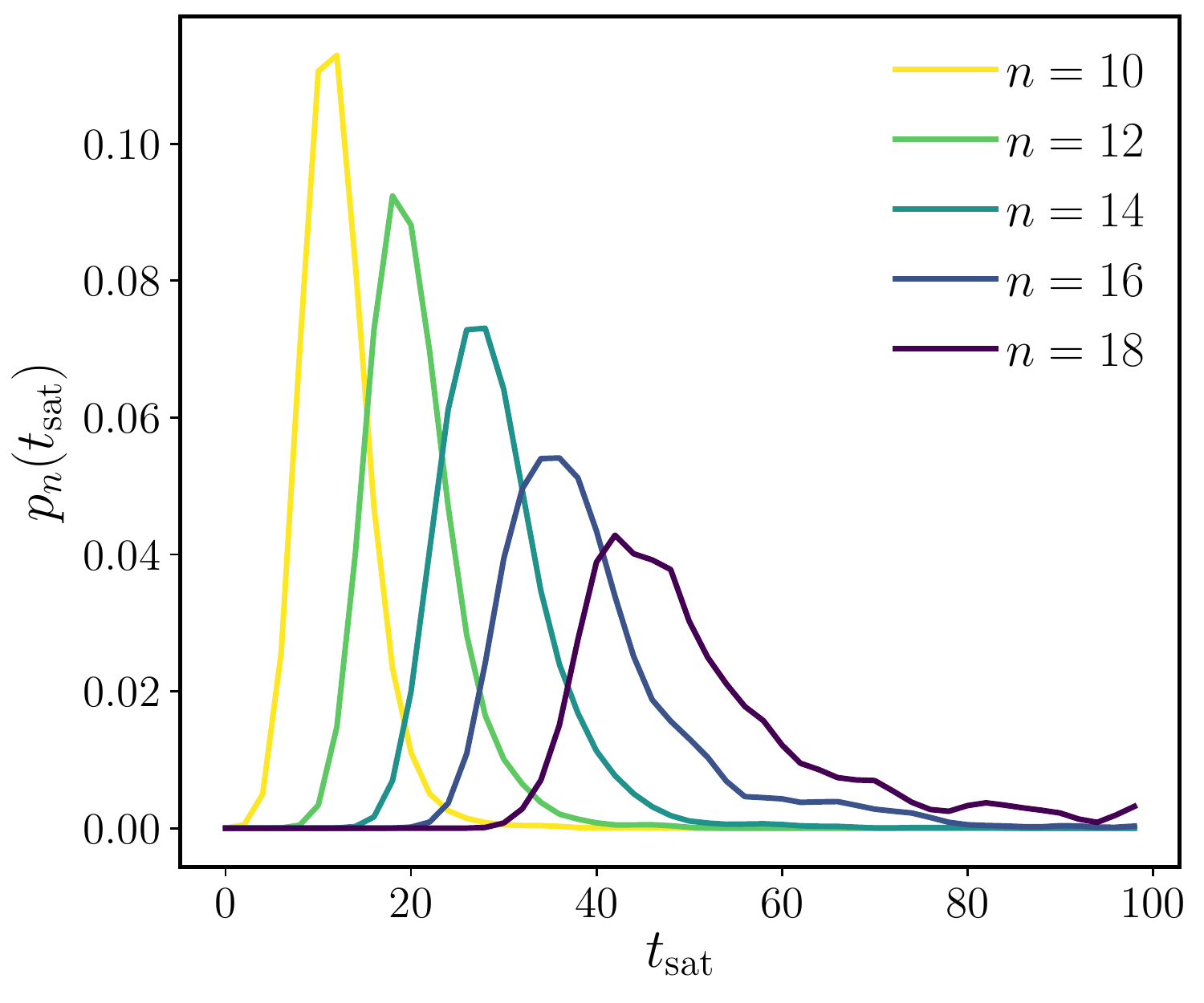}
    \caption{Distributions of the saturation time at the NEE point (\( \Gamma = 0.4 \)). The data was smoothed through integration against a Gaussian kernel of width \( \sigma = 2 \).}
\end{figure}

\begin{figure}
  \centering
    \label{fig:right_tail_cdf}
    \includegraphics[width=0.5\textwidth]{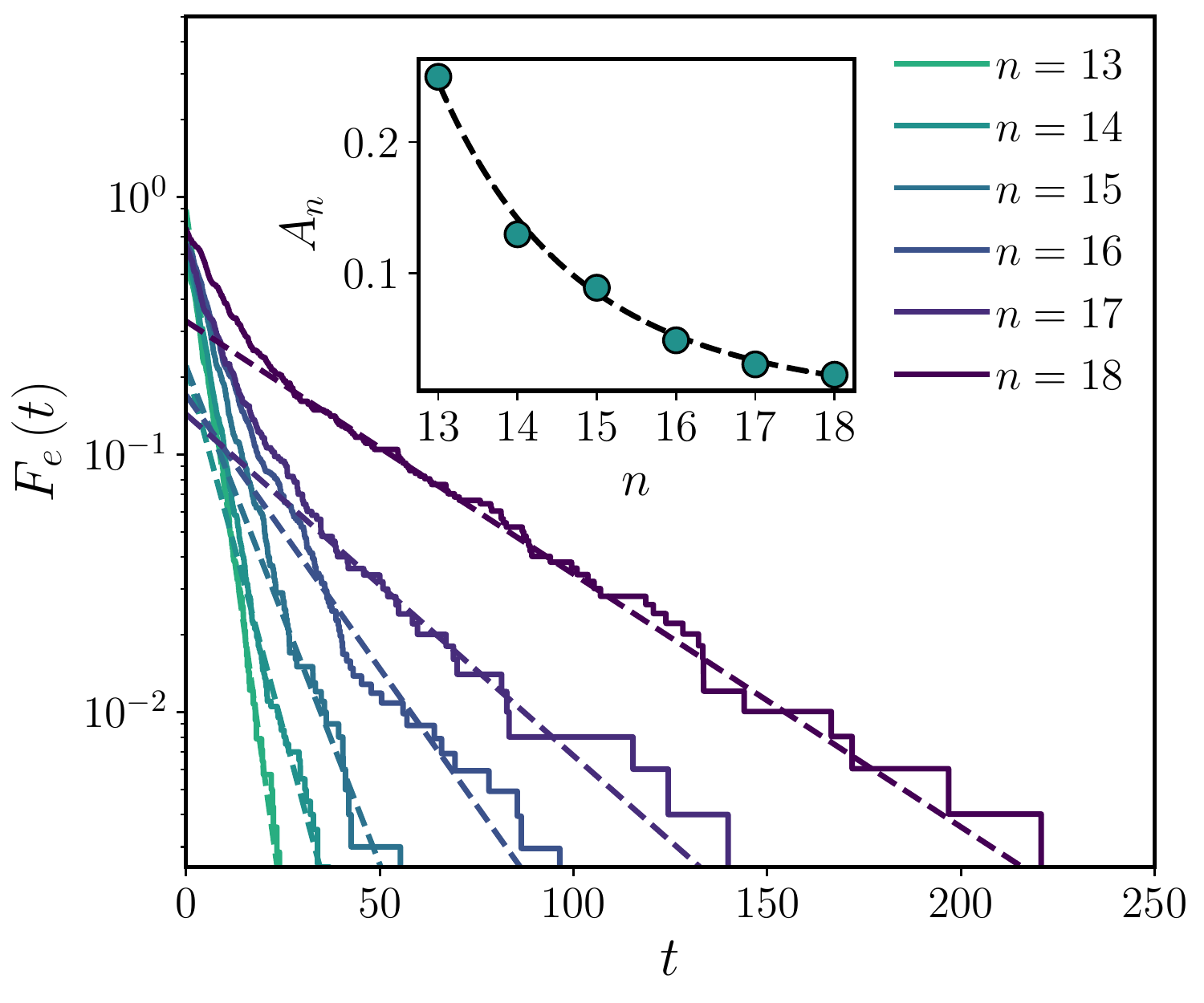}
    \caption{Right tails of the distributions in Fig.~\ref{fig:teq_histograms}, as described by the empirical CCDF, Eq.~\eqref{eq:eccdf}. Dashed lines show the exponential fits, Eq.~\eqref{eq:ccdf_fit}. \emph{Inset:} exponent \( A_n \), as defined in Eq.~\eqref{eq:ccdf_fit}, describing the decay rate of the distribution of \( t_\sat \) at large values of its argument. The dashed line represents the power-law fit \( A_n \sim an^{-b} \), with \( b = 7.4(2) \).}
\end{figure}

We now study the distribution of saturation times for the PT process. Fig.~\ref{fig:teq_histograms} shows the histograms of the saturation times separated by system size $n$. Note that they exhibit a peak with a long right tail. We can use a (truncated) complementary cumulative distribution function (CCDF)
$$
    F(t) \equiv \int_{t_{\mathrm{peak}}+t}^{\infty} p(t_\sat)\,\mathrm{d}t_\sat
$$
to study the right tail of the distributions. $t_{\mathrm{peak}}$ is the position of the peak of $p(t_\sat)$ estimated through a Gaussian smoothing of the data. We approximate $F(t)$ using the \emph{empirical} CCDF $F_e(t)$ obtained from the raw data $\{t_\sat^{(i)}\}_{i=1}^N$
\begin{equation}\label{eq:eccdf}
    F_{e}(t) = \frac{1}{N}\sum_{i=1}^N \Theta\Big(t_\sat^{(i)}-t-t_\mathrm{peak}\Big)
\end{equation}
where $\Theta$ is the Heaviside step function. We find that $F_e(t)$ is well-approximated by a two-parameter exponential functional form (Fig.~\ref{fig:right_tail_cdf})
\begin{equation}\label{eq:ccdf_fit}
    F_n(t_\sat) \approx \exp\Big(-A_nt_\sat-B_n\Big) 
\end{equation}
where (Fig.~\ref{fig:right_tail_cdf}, inset)
$$
A_n \sim an^{-b}
$$
so
$$
    F_n(t_\sat) \sim \exp\Big(-a\frac{t_\sat}{n^{b}}\Big) 
$$
which means that the probability density $p_n(t_\sat)$ for large values of $t_\sat$ is
$$
p_n(t_\sat) \sim an^{-b}\exp\Big(-a\frac{t_\sat}{n^{b}}\Big)
$$
that is, even though the probability distributions $p_n$ have thin (\ie exponentially-decaying) right tails at all finite sizes $n$, their decay coefficients get increasingly smaller with $n$ so the tails get fatter with increasing $n$.

\begin{figure}
  \centering
    \label{fig:teq_boxplot}
    \includegraphics[width=0.5\textwidth]{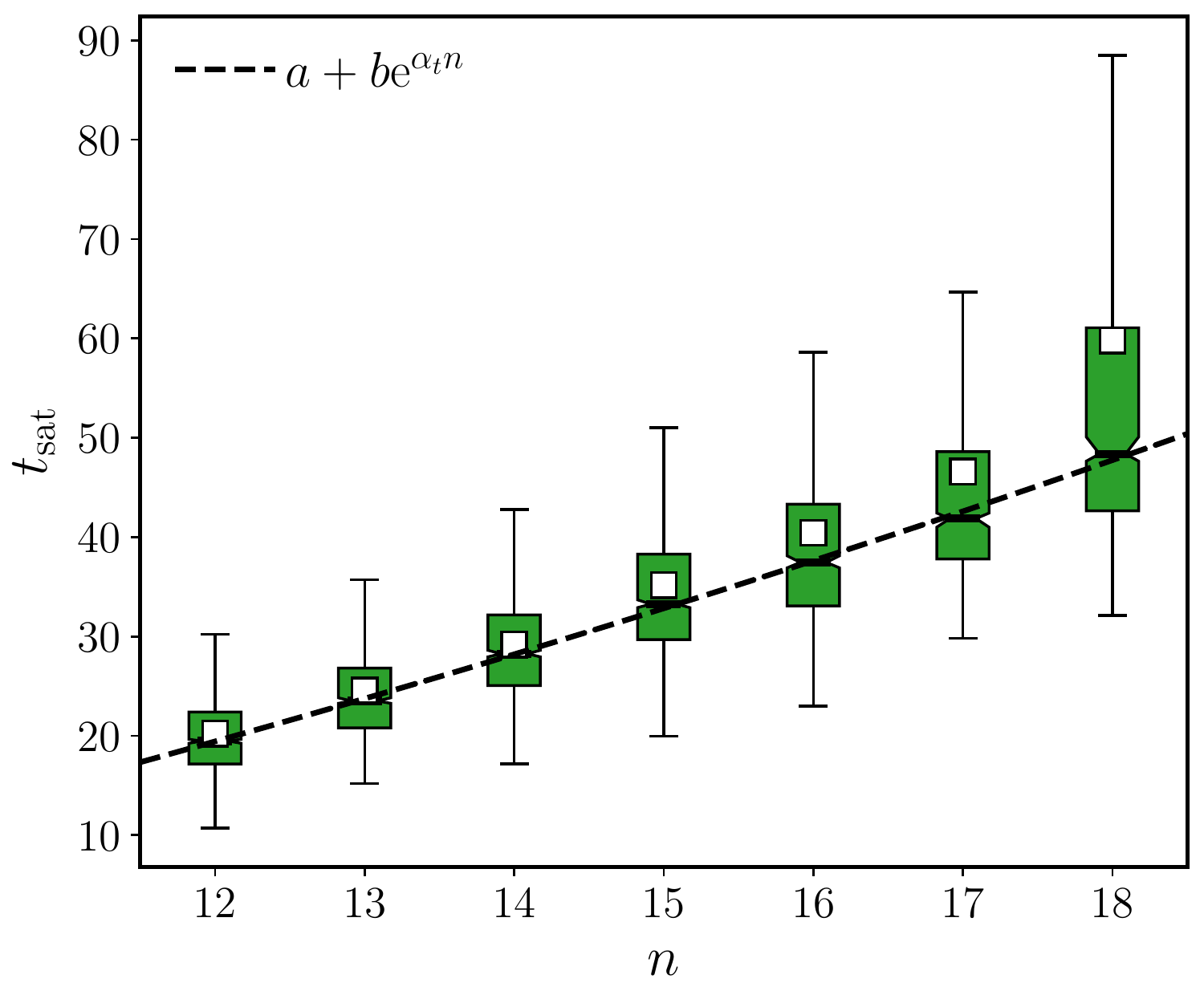}
    \caption{Box plot of the distributions in Fig.~\ref{fig:teq_histograms}. Boxes subsume the middle half of the data. The black bar in the middle of each box is the median, with its bootstrapped error denoted by notches in the box. White squares represent the mean values. Whiskers extend from the smallest datapoint above \( x_\mathrm{low} = q_1 - \frac{3}{2}(q_3-q_1) \) to the largest one below \( x_\mathrm{high} = q_3 + \frac{3}{2}(q_3-q_1) \), where \( q_1 \) and \( q_3 \) are the first and third quartile, respectively. The dashed line depicts a shifted exponential fit of the median times, with \( \alpha_t = 0.04(2) \).}
\end{figure}

\section{PT quality: spread and spillage}\label{sec:algorithm}
In this Section we start to analyse the PT dynamics as a solution-mining algorithm. In order to benchmark its performance we need to study quantitatively how the time-evolved wavefunction populates the target subspace $\Omega_n$. In an ideal situation, one would like that 1) the wavefunction $\ket{\psi(t_{\infty})}$ should be completely contained in the target subspace, and 2) it should be uniformly spread, that is
\begin{equation} \label{eq:d_exponent}
\lvert \psi_i(t_{\infty}) \rvert^2 = \begin{cases}
1/\lvert \Omega_n \rvert & \text{if $i \in \Omega_n$}\\
0 & \text{otherwise,}
\end{cases}
\end{equation}
as this would provide an effective way of finding all target solutions efficiently by repeatedly sampling the distribution $\lvert \psi_i \rvert^2$ (``fair sampling'', see \eg \cite{PhysRevE.99.063314}). This is of course an ideal limit, and is never achieved in practice. Nevertheless, we can use the displacement from this ideal case as a yardstick for PT success. We will first consider the following two quantities.

\begin{enumerate}
\item the total probability $L_\Omega[\psi] = \lVert P^{\bot}_{\Omega}\ket{\psi(t)} \rVert^2$ of finding a state outside of the target subspace (\emph{amplitude spillage}), and
\item the degree of uniformity of the wavefunction's intensity distribution with respect to the Hamming distance, which we measure by means of a functional \( U_\Omega[\psi] \) described in detail in Appendix~\ref{app:U}. This functional satisfies \( U_0 \le U_\Omega[\psi] \le 1 \), taking its minimal value \( U_0 \approx 0.6065 \) when the restriction of \( |\psi_i|^2 \) to \( \Omega \) is completely uniform and its maximal value 1 when it is atomic.
\end{enumerate}

\begin{figure}
  \centering
    \includegraphics[width=0.4\textwidth]{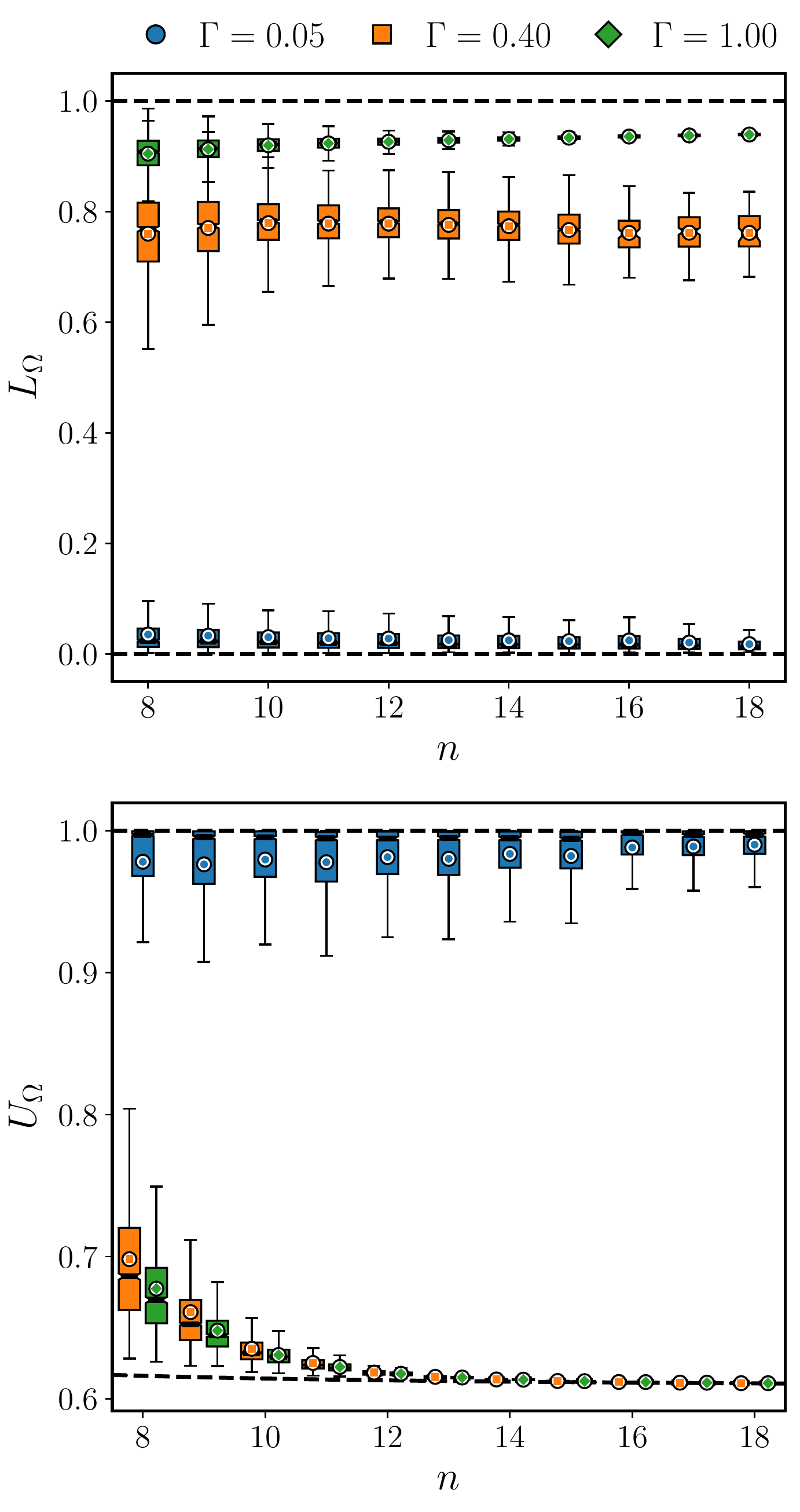} \label{fig:spread_spillage}
    \caption{\emph{Top:} distribution of the \( L_\Omega \) (``spillage'') functional, discriminating ergodic regimes (\( L_\Omega \approx 1 \)) from nonergodic ones (\( L_\Omega < 1 \)). \emph{Bottom:} distribution of the \( U_\Omega \) functional, discriminating localized regimes (\( U_\Omega \approx 1 \)) from extended ones (\( U_\Omega \approx ((1+e^{-1/n})/2)^n \); cf.\ Appendix~\ref{app:U}). For an explanation of the box plot, see Fig.~\ref{fig:teq_boxplot}.}
\end{figure}

In Fig.~\ref{fig:spread_spillage} we show the behaviour of these two quantities in three points of the phase diagram that we take as exemplary points for the localized ($\Gamma=0.05$), NEE ($\Gamma=0.4$) and ergodic behaviour ($\Gamma=1$). All points are associated to the same energy density $\epsilon=0.27$. As a first sanity check, note that
\begin{enumerate}
\item the spread of the wavefunction component in the target subspace (as measured by the \( U_\Omega \) functional) is essentially the same in the NEE and in the ergodic case.
\item the localized case has the least amount of spillage but is also the most inhomogeneously spread.
\end{enumerate}

\begin{figure}
  \centering
    \label{fig:psiwo}
    \includegraphics[width=0.5\textwidth]{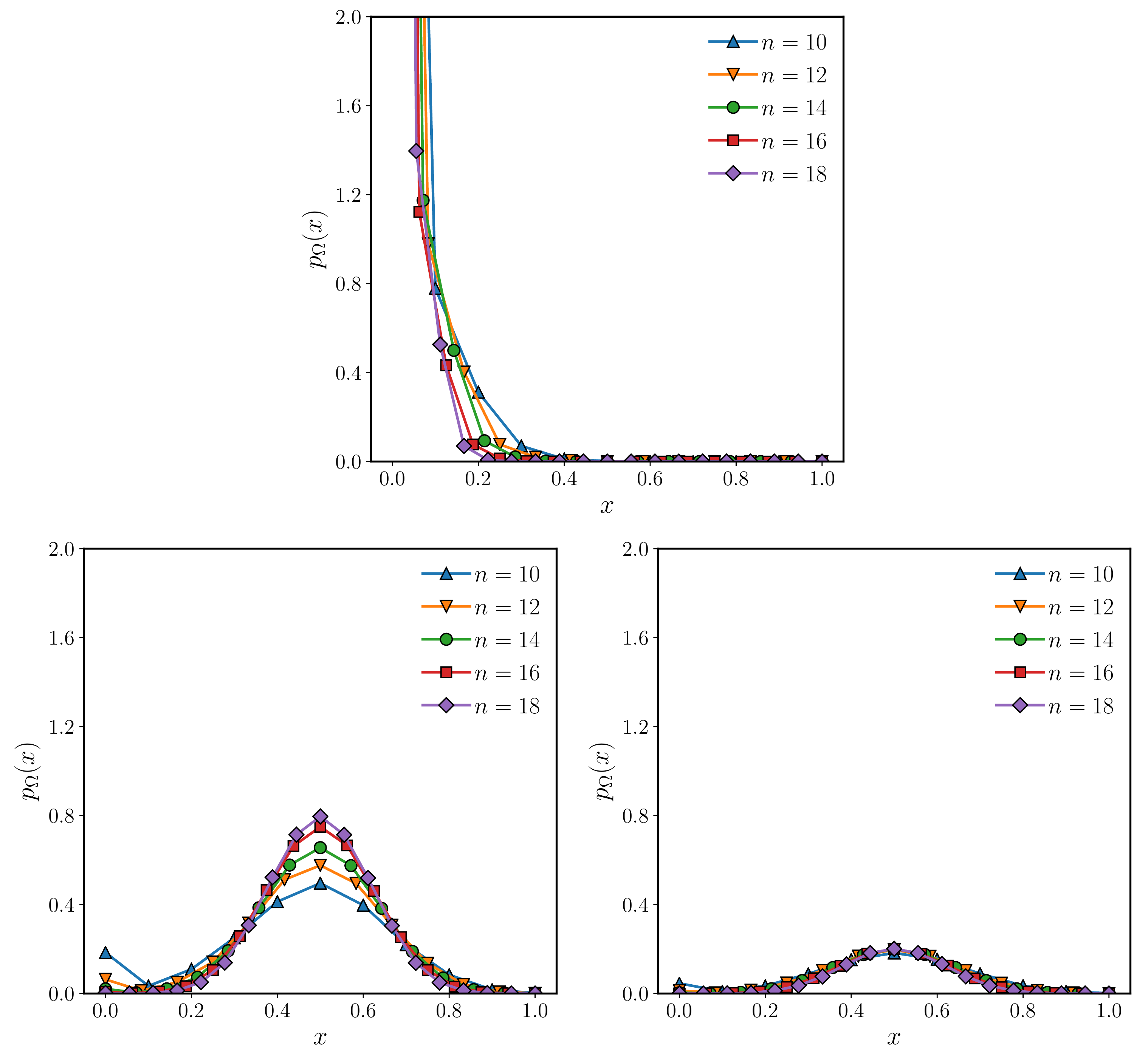}
    \caption{Probability distributions of finding a resonance via PT as a function of its fractional distance \( x = d/n \) from the initial state, at energy density \( \epsilon = 0.27 \) and transverse field \( \Gamma = 0.05 \) (top), 0.40 (bottom left) and 1.00 (bottom right).}
\end{figure}

In order to see why this is the case we plot the total probability inside of the target subspace $p_\Omega = \sum_{z\in\Omega} \lvert \langle z | \psi\rangle \rvert^2$, resolved by the fractional Hamming distance $x = \operatorname{dist}(z,z_0)/n$ from the initial state $z_0$ (Fig.~\ref{fig:psiwo}). Some comments about these results.
\begin{enumerate}
\item In the localized case, the amplitude is more and more concentrated on the initial state $z_0$. The PT protocol in this phase will not be able to efficiently find target states that are more than $d = xn$ spin flips away from $z_0$, for large enough $0 < x < 1$.
\item In the ergodic case we see a roughly Gaussian distribution of the amplitudes over the fractional Hamming distance. This is an effect of the geometry of the Boolean hypercube (most states are $n/2$ spin flips away from the initial state $z_0$). Note that the distributions seem to be essentially independent of $n$ (within the statistical errors due to the finite sampling of the disorder). 
\item In the NEE case we see the same overall Gaussian shape of the distributions, but in this case they are larger and they flow favourably with $n$. This is due to the fact that more amplitude is contained inside the target subspace than in the ergodic case.
\end{enumerate}

\section{Quantum advantage}
Thus far we have learned some facts about the PT protocol, but its application to quantum computing is practically relevant only insofar as it can be cast as a quantum algorithm that is able to outperform all classical algorithms at some specific computational task, a situation described as \emph{quantum advantage} (or \emph{speedup} if time is the relevant quantity). In order to show quantum advantage in the energy matching problem described in Section \ref{sec:introduction}, one would like to compare the running time of the PT algorithm against an optimal classical algorithm designed to find all states at a given energy. This is generally a significantly understudied problem. Some very specific models have provably efficient algorithms that sample the Gibbs distribution for a given temperature (\eg the planar Ising spin glass \cite{PhysRevE.80.046708}), so that if one is able to compute the Legendre transform between energy $E$ and temperature $T$ then in the large $n$ limit one could obtain all states at a given energy by sampling the Gibbs state of the associated temperature $T_E$. Failing that, one is reduced to using a general-purpose stochastic local search algorithm such as some form of simulated annealing. 

Unfortunately, such comparison is not very useful in the case of the REM due to its trap-model dynamics: its local dynamics is made up of a sequence of thermally-activated events where the system is excited out of a low-energy state to the flat plateau (the classical states with $\epsilon=0$), followed by a random walk on the plateau until it falls again into another (random) low-energy state with a random energy $E$. If $E$ lies inside of the target energy shell, then we have found a target state of the energy matching problem and we are done. If not, we need to wait for the system to be thermally excited out of this state and the search process starts anew. Even assuming one can speed up the excitation events so that they only take $O(1)$ time, there seems to be no way to avoid having to do a random walk on the plateau in order to find a new low-energy state. Due to the flatness of the plateau and the random relative positions between states of non-typical energy density, this stochastic local search seems to be intuitively equivalent to performing a global random search.

Therefore we use a different benchmark metric. For a given realization of the REM Hamiltonian $H_p$, we define two ``oracles'', $\mathcal{O}_{\mathrm{qu}}$ and  $\mathcal{O}_{\mathrm{cl}}$, that take as input an initial state $z_0$ and attempt to produce a target state $z^\prime \in \Omega_n$. The first is a quantum oracle that applies a PT evolution to $\ket{z_0}$ for a time $t = t_{\sat}$, and $\Gamma = O(1)$, and then measures the final state $\ket{\psi(t)}$ in the computational basis. The second oracle is a classical procedure that discards the initial state $z_0$ and simply samples a new state $z \in \{0,1\}^n$ uniformly at random, with equal probability $1/2^n$. The probability of success \emph{for one call} of each oracle is
\begin{eqnarray*}
P_{\mathrm{qu}} &=& \sum_{z \in \Omega_n} \lvert \langle z | \psi(t) \rangle \rvert^2\\
P_{\mathrm{cl}} &=& \lvert \Omega_n \rvert/2^n.
\end{eqnarray*}
We define the ``gain'' $G \equiv P_{\mathrm{qu}}/P_{\mathrm{cl}}$ as the ratio of these two probabilities. The gain is a random variable due its dependence on 1) the problem instance $H_p$, and 2) the choice of initial state $z_0$. 

\begin{figure}
  \centering
    \label{fig:fss_gain}
    \includegraphics[width=0.5\textwidth]{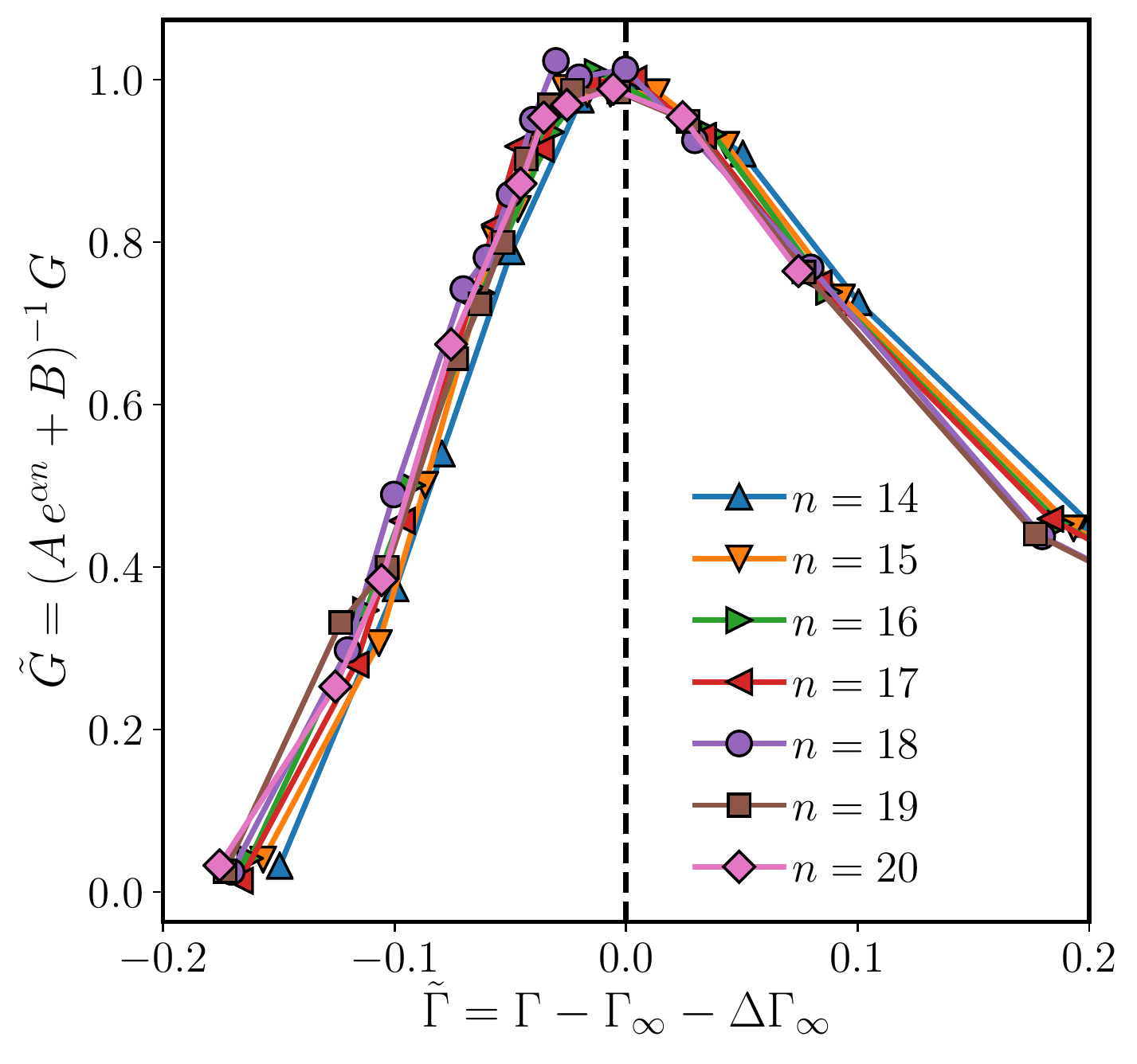}
    \caption{Curve collapse of the median gain for system sizes \( n = 14 \) through 20.}
\end{figure}

For each size $n = 8, \ldots, 20$ we sampled a number of disorder realizations of the REM Hamiltonian $H_p$ and for each of them we selected the state $z_0$ with an energy closest to $E = 0.27n$. The probability $P_{\mathrm{cl}}$ can be computed by simple inspection of the random energies in the realization $H_p$, while to compute $P_{\mathrm{qu}}$ we simulated the PT protocol numerically.
We performed a finite-size scaling of the data thus obtained (Fig.~\ref{fig:fss_gain}) using the Ansatz
$$
    G(n,\Gamma) = (Ae^{\alpha n} + B) g\big(\tilde{\Gamma}(n)\big)
$$
where the effective $\tilde{\Gamma}(n)$ is given by
\begin{equation}\label{eq:peak_ansatz}
\tilde{\Gamma}(n) \equiv \Gamma - \Gamma_\infty + \frac{C}{n^{1/\nu}}.
\end{equation}
Given the available data, fitting three parameters results in a poor estimate for the uncertainties on \( C \) and \( \nu \). Therefore, we prefer to fix one of them while keeping the other one free. Observing that the gain peaks approximately in correspondence of the Anderson transition, we chose to utilize the same value of \( \nu \) describing the finite-size shift of the mobility edge, which is known to be $0.3 \leq \nu \leq 0.5$ \cite{Baldwin2016}. We use $\nu = 0.4$ so that $1/\nu=2.5$ \footnote{we remark that even if we leave $\nu$ as a free parameter in the fit, we obtain a value of approximately $1/\nu = 2.53$.}. We obtain the following values for the fitting parameters:
\begin{align*}
\alpha &= 0.084 \pm 0.005\\
A &= 5.3 \pm 0.8\\
B &= -2.7 \pm 1.3\\
C &= 25.85 \pm 1.2\\
\Gamma_\infty &= 0.242 \pm 0.001.
\end{align*}
This means that
\begin{enumerate}
    \item the median gain peaks at a size-dependent value of $\Gamma_{\mathrm{peak}}(n) $ that for $n\rightarrow \infty$ is in quite good agreement with the estimate of the Anderson transition's mobility edge ($\Gamma_c \approx 0.27$ following the forward scattering approximation). The thermodynamic-limit position of the peak $\Gamma_{\infty}$ is shifted by a finite-size effect $\Delta\Gamma_{\infty}(n) = -Cn^{-1/\nu}$. This shift was already observed in Ref.~\cite{Baldwin2016} in the finite-size scaling analysis of the mobility edge of the QREM using the forward-scattering approximation, as well as other quantum phase transitions in disordered transverse-field models \cite{Mossi_2017}.
    \item the height of the peak scales like $G_{\mathrm{peak}} \sim e^{\alpha n}$ with $\alpha > 0$. This is a query-complexity asymptotic advantage of the quantum PT oracle over the random search oracle.
\end{enumerate}

\begin{figure}
  \centering
    \label{fig:gain_peak}
    \includegraphics[width=0.47\textwidth]{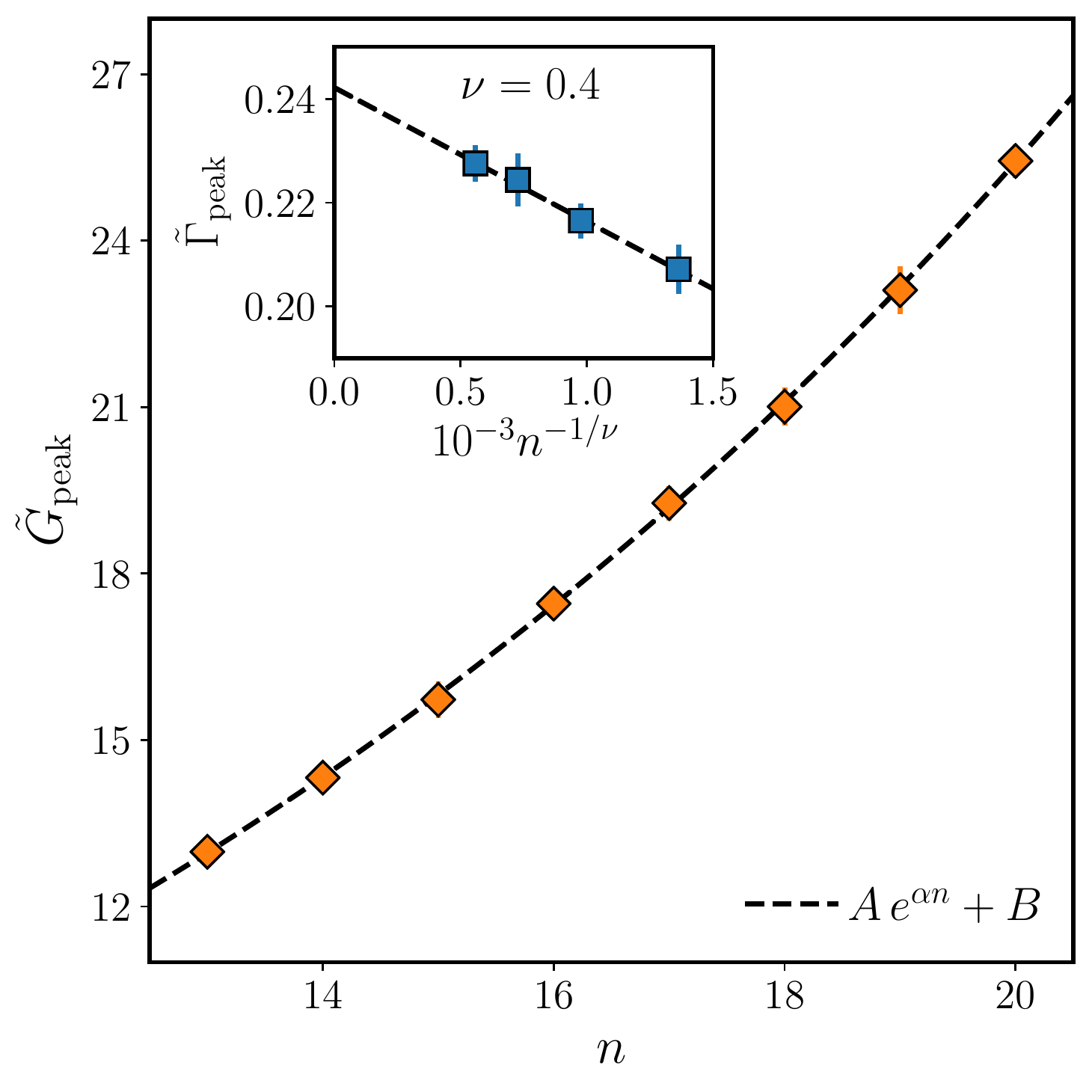}
    \caption{Peak median gain for \( n = 13, 14, \dotsc, 20 \), fitted with a shifted exponential \( A\,e^{\alpha n} + B \) with parameters \( A = 5.3(8), \alpha = 0.084(5), B = -27(13)\cdot 10^{-1} \). \emph{Inset:} position of the peak median gain for \( n = 14, 16, 18, 20 \). The scaling Ansatz Eq.~\eqref{eq:peak_ansatz} suggests a value \( \tilde{\Gamma}_\mathrm{peak} \approx 0.24 \) in the thermodynamic limit.}
\end{figure}

\subsection*{Non-oracle Advantage}
In order to assess the absolute (\ie non-oracular) advantage of the PT protocol over stochastic random search one has to compute the full runtime of both algorithms by multiplying the number of oracle calls by the time required to implement a single oracle call. Equivalently, one can study the quantity
\begin{equation}\label{eq:k_advantage}
K = \frac{P_\qu}{t_\qu} \frac{t_\cl}{P_\cl} = G \frac{t_\cl}{t_\qu},
\end{equation}
where $t_\qu,t_\cl$ are the time needed to implement one call to the quantum and classical oracles (respectively). For the classical oracle, a random search requires a time $t_\cl = O(n)$ as one only needs to generate $n$ random bits (linear time for a classical probabilistic Turing Machine). For the quantum oracle, $t_\qu$ is the saturation time $t_\sat$ we studied in Subsection \ref{sec:dynamics}. Therefore $K(n) = n G(n)/t_\qu(n)$. We consider a setup where, for fixed $\Gamma=0.4$ (that represents a good choice of parameter setting without being optimal) we evolve the initial state according to the PT protocol for a time $t=t_\sat(n)$ and then we sample the wavefunction in the computational basis. Quantum advantage is decided by the competition of the gain $G$ and the saturation time $t_\sat$. 

Unfortunately, the data at finite sizes $n=10-20$ show only a weak dependence on $n$ so that the functional form of these two quantities is hard do extract conclusively. On physical grounds we expect exponential scaling of both $G$ and $t_{\sat}$, so we fit through functions of the form
\begin{equation}\label{eq:fit_g_t}
f(n) = a e^{\alpha n} + b, \quad \text{$a,b,\alpha \in \mathbb{R}$}.
\end{equation}

The values of the $\alpha$ exponents control the rate of divergence of the leading term of these quantities in the limit $n\rightarrow \infty$. We use the notation $\alpha_t$ and $\alpha_g$ for the time and gain exponent, respectively. We obtain the values

\begin{eqnarray}
\alpha_g &=& 0.09 \pm 0.04 \\
\alpha_t &=& 0.04 \pm 0.02
\end{eqnarray}

Thus we have the following approximate scaling
$$
K = \frac{n\,G(n)}{t_\sat(n)} \sim \exp\Big((\alpha_g - \alpha_t)n\Big) \approx \exp\Big((0.05 \pm 0.04)n\Big).
$$
While this would imply an asymptotic speedup in the $n\rightarrow \infty$, the prefactor of $n$ at the exponent is very small and we believe that the most reasonable interpretation of these results is that the PT protocol and random search scale approximately in the same way, at least at the sizes we have been able to study.

\begin{figure}
  \centering
    \includegraphics[width=0.5\textwidth]{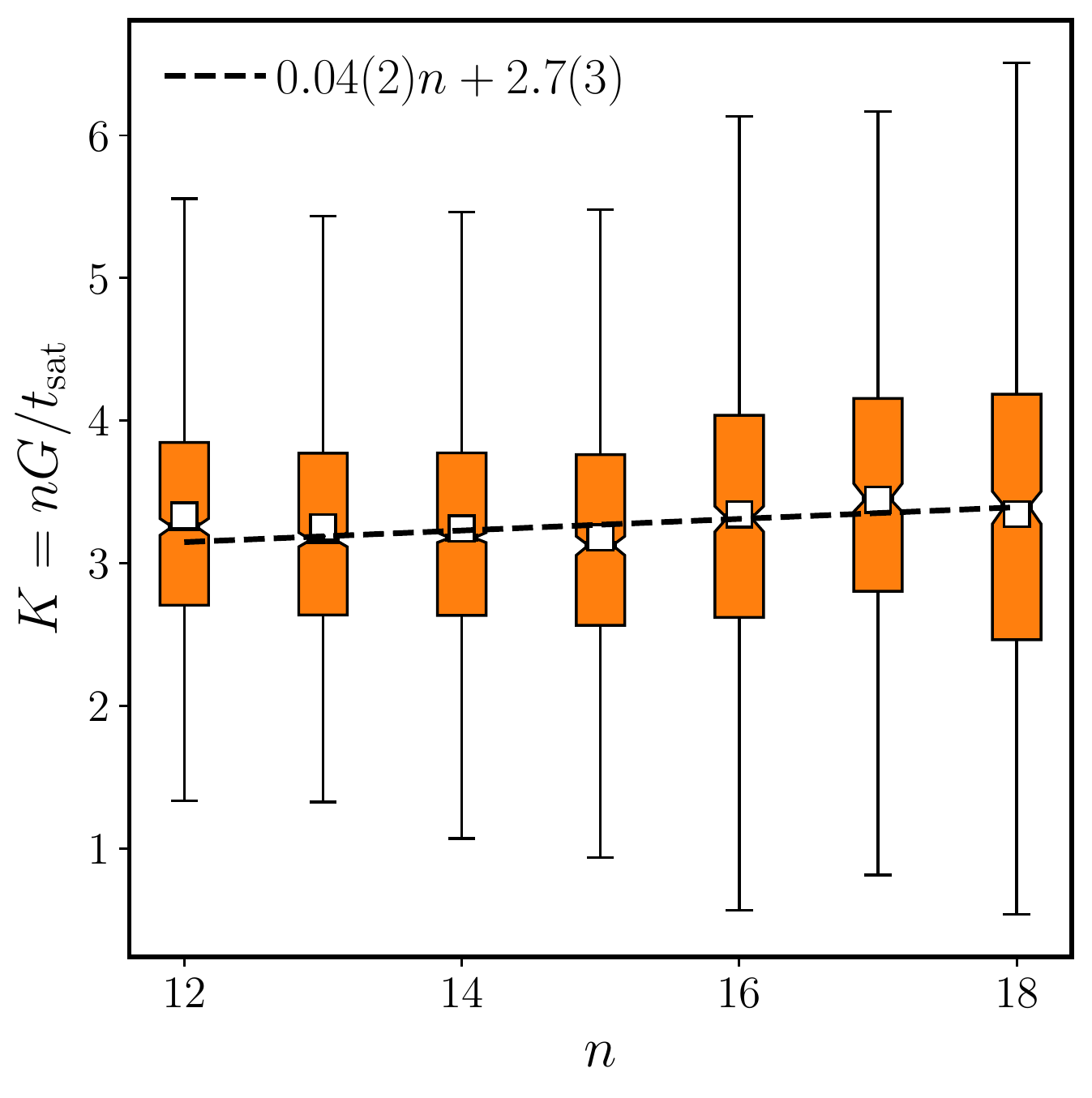} \label{fig:K_plot}
    \caption{Distributions of the value of \( K \), as defined in Eq.~\eqref{eq:k_advantage}. For an explanation of the box plot, see Fig.~\ref{fig:teq_boxplot}.}
\end{figure}

Indeed, in case of a very small (or even zero) value for $\alpha_g-\alpha_t$, the linear factor that comes from $t_{cl}$ in Eq.~\eqref{eq:k_advantage} might affect the behaviour of $K$ at small $n$. However, this does not seem to make much of a difference for the system sizes we can observe: if we plot the median values of the $K$ value directly (instead of fitting the exponential scaling of $G$ and $t_{\sat}$ separately) then the corrections to the exponential terms (coming from the parameters $a,b$ in Eq.~\eqref{eq:fit_g_t} and the linear term $n=t_{cl}$) still cancel out any significant difference between $G$ and $t_{\sat}$, and the slope of $K$ turns out to be extremely small (see Fig.~\ref{fig:K_plot}) 

Our conclusion is that in the QREM, the PT protocol and random search scale approximately in the same way, for problem sizes $n=12-20$. The small curvature of both the time data and the gain data suggests that going to moderately larger sizes, as in some of the largest numerical analysis currently available in the many-body localization literature \cite{10.21468/SciPostPhys.5.5.045}, is unlikely to improve our analysis very much. Any eventual asymptotic advantage would be significant only for significantly larger sizes. 

\section{Conclusions}\label{sec:conclusions}
In this work we have studied numerically (up to system sizes of $n=20$ quantum spins) the population transfer dynamics in the QREM with the goal of assessing a possible application in quantum computing. 

For a horizontal line in the $(\epsilon,\Gamma)$ phase diagram of the model, corresponding to fixed energy density $\epsilon=0.27$, we computed the scaling of the support size of both the energy eigenfunctions, and the wavefunctions obtained at the end of the delocalization process induced by the system's coherent dynamics applied to a localized initial state. For increasing values of $\Gamma$, we observed in both cases a continuous crossover of the scaling dimension $D$ from a localized ($D=0$) towards an ergodic regime ($D=1$), which indicates the presence of a non-ergodic extended phase.

We further studied the timescale $t_\sat$ required for the PT delocalization process to complete, as a function of the system size $n$. Over the disorder, this time follows a unimodal distribution with a long right tail that gets fatter with increasing $n$. The median value of $t_\sat$ shows an exponential dependence of on the system size, $t_\sat\sim \exp(\alpha n)$.

We computed the probabilities $P_\qu, P_\cl$ of obtaining a state in the target microcanonical shell (containing the states with the same energy density as the initial state), using respectively the saturated PT dynamics for a given value of $\Gamma$, and random search. The ratio $G = P_\qu/P_\cl$ (``gain'') of these probabilities corresponds to a query-complexity comparison between a single use of a black box that performs the PT protocol to completion, and then samples the resulting wavefunction in the $\sigma^z$ basis, and a single use of a black box that samples classical states uniformly at random. We observe that the gain grows with $n$ in the regime of sizes that is accessible to our numerics for all values of $\Gamma$ we considered --- \ie the quantum PT oracle increasingly outperforms random search --- with a peak that appears in proximity of the critical point of the Anderson transition.

However, if we compare the actual (\ie non-oracular) runtime of the two algorithms by dividing the probability of success of the black boxes by the time required to implement one call to the black box, we find that the scaling of the gain $P_\qu/P_\cl$ and the scaling of the inverse times ratio $t_\cl/t_\qu$ roughly cancel out: at the system sizes considered in this work, PT and random search seem to scale with $n$ in approximately in the same way.

We believe that the final considerations we can extract out of the above results are twofold, one concerning the choice of parameters for the PT protocol and another concerning the overall efficiency of the PT protocol as a quantum algorithm for energy matching. For what concerns the parameter setting, the fact that the optimal choice of $\Gamma$ seems to coincide with the critical point of the localization/delocalization transition of the QREM suggests that the parameter setting problem for $\Gamma$ can be reduced to finding the mobility edge for the Anderson transition. This is a problem that has undergone extensive studies in the literature on many-body localization. Our conjecture is that this optimality result should generically hold true for any other combinatorial optimization problem to which PT can be applied. For the second PT parameter, the final evolution time $t_{\mathrm{fin}}$, the situation is less clear. The choice of an optimal time is hard to do on a case-by-case basis, as we found no obvious property of a REM instance that correlates with its saturation time $t_{\sat}$. At this point we are only able to suggest that exploratory runs should be made using randomly generated instances in order to estimate a fixed percentile of the distribution of $t_{\sat}$ (\eg its median), so that one is guaranteed that for that choice of $t_{\mathrm{fin}}$, a finite fraction of instances will have reached saturation and the optimal performance of PT protocol will be reached with finite probability. Clearly, more work will be necessary in order to elucidate this point before PT will be ready for practical applications.

The second point concerns the efficiency of PT for energy matching: according to our findings, PT does not show significant quantum advantage on the QREM 1) at the sizes that are likely to be accessible to near-term quantum technologies, 2) in the entire NEE phase. This leaves a few open scenarios that we are unable to resolve at the moment:
\begin{enumerate}
    \item PT never exhibits asymptotic advantage in the QREM, for any energy density whatsoever. This would confirm the asymptotic results obtained by the authors of Ref.~\cite{Baldwin2018} by keeping the leading-order term of the Schrieffer--Wolff perturbation theory in the small parameter $\Gamma$.
    \item PT shows asymptotic advantage but only for a restricted interval of energy densities in the NEE phase. In general, one would expect these energies to be neither too close to the edges of the spectrum nor too close to the center. Currently there is not to our knowledge a way of detecting these ``good'' energies from finite size data (all approaches consider asymptotic studies in $n$), except for exhaustively trying them all.
    \item PT has a (perhaps even ``almost quadratic'', similar to Ref.~\cite{Smelyanskiy2020}) quantum advantage in the QREM, but the system sizes we can access are simply too pre-asymptotic: the scaling we are seeing at these system sizes is too different from the scaling that emerges in the $n \rightarrow \infty$ limit.
\end{enumerate}
We believe that this question will not be resolved at the sizes $n = 20-30$ that seem to be the current limit of numerical methods, or will likely be achieved in the next few years. In order to decide among the scenarios above it seems reasonable to us that better numerical methods need to be developed in order to approximate the effective coherent dynamics of the QREM in a fixed energy shell so that much larger system sizes can be investigated. Downfolding techniques like the ones developed in \cite{Smelyanskiy2020} are likely to prove useful, when paired with a robust way of estimating the error of the effective propagator for given time $t$ and finite size $n$.

One interesting future line of research involves the experimental realization of the PT protocol on quantum hardware. While unfeasible in our case due to the non-local nature of the REM Hamiltonian, this is in principle achievable on other finite-connectivity combinatorial models. Indeed, beyond the QREM, we believe it would be interesting to consider studying the PT dynamics in models with correlated disorder (\eg $k$-SAT). In these other cases the expectation is that the scaling of the gain will be worse compared to the QREM, as the comparison must be made with classical algorithms or heuristics that can exploit the correlated energy landscape and therefore outperform random search. However, the inverse time ratio ${t_\cl}/{t_\qu}$ in Eq.~\eqref{eq:k_advantage} is likely to scale more favourably: the runtime $t_\cl$ for these probabilistic algorithms to produce one candidate solution for the energy matching problem (\ie a purported state of the correct target energy) will likely grow exponentially with the system size --- as is expected to happen for essentially any ``hard'' optimization problem --- unlike what happens in random search, where one such attempt only requires the generation of $n$ random bits (which is the reason for the $t_\cl = O(n)$ factor in our case).
These local models are particularly attractive as they can be simulated directly on quantum hardware for much larger system sizes than the ones accessible to numeric methods, possibly solving the question of the asymptotic behaviour of PT. Digital quantum computers are limited by the fact that they can execute programs only in the form of quantum circuits: the simulation of continuous-time processes on digital quantum hardware almost always requires the discretization of the real-time unitary propagator \eg via a Trotter decomposition, even in case of time-independent Hamiltonians such as the ones used in the PT protocol. This significantly limits the final time that can be reached by the simulation, as the elimination of Trotter errors necessitates the use of quantum circuits of non-negligible depth. Analog quantum computing technology, on the other hand, natively implements continuous-time quantum evolution, and at this point seems to be tantalizingly close to being able to simulate the PT protocol. However, currently available quantum annealers such as the D-Wave machine were not designed with PT in mind. Consequently, their annealing schedule --- while more complicated than PT --- is currently too rigid and cannot accomodate long intervals where the transverse field is kept constant, that are instead a necessary element of PT. We believe that the developement of high-coherence analog quantum computing hardware with a broader range of applicability will greatly benefit the study of population transfer protocols for energy matching.
Finally, there remains the unexplored option of using the PT dynamics for a hybrid ``PT-assisted stochastic search'', where quantum PT moves are alternated with classical stocastic search (\eg some form of simulated annealing) either for purpose of energy matching or for combinatorial optimization, as mentioned \eg in \cite{1807.04792}. This is an interesting approach that might boost the performance of these classical algorithms in specific conditions.

\section{Acknowledgements}\label{sec:acknowledgements}
The authors would like to thank Vadim Smelyanskiy and Boris Altshuler for useful discussions, and Antonello Scardicchio and Eleanor Rieffel for helpful suggestions concerning the first draft of this paper. T.~P. would like to thank the Institute for the Theory of Quantum Technologies (TQT) of Trieste for support, the QuAIL group at NASA Ames and the Universities Space Research Association (USRA) for the kind hospitality and support while part of this work has been done. We are grateful for support from NASA Ames Research Center, the AFRL Information Directorate under grant F4HBKC4162G001, and the Office of the Director of National Intelligence (ODNI) and the Intelligence Advanced Research Projects Activity (IARPA), via IAA 145483. The views and conclusions contained herein are those of the authors and should not be interpreted as necessarily representing the official policies or endorsements, either expressed or implied, of ODNI, IARPA, AFRL, or the U.S. Government. The U.S. Government is authorized to reproduce and distribute reprints for Governmental purpose notwithstanding any copyright annotation thereon.
\bibliography{qrem}

\begin{thebibliography}{35}%
\makeatletter
\providecommand \@ifxundefined [1]{%
 \@ifx{#1\undefined}
}%
\providecommand \@ifnum [1]{%
 \ifnum #1\expandafter \@firstoftwo
 \else \expandafter \@secondoftwo
 \fi
}%
\providecommand \@ifx [1]{%
 \ifx #1\expandafter \@firstoftwo
 \else \expandafter \@secondoftwo
 \fi
}%
\providecommand \natexlab [1]{#1}%
\providecommand \enquote  [1]{``#1''}%
\providecommand \bibnamefont  [1]{#1}%
\providecommand \bibfnamefont [1]{#1}%
\providecommand \citenamefont [1]{#1}%
\providecommand \href@noop [0]{\@secondoftwo}%
\providecommand \href [0]{\begingroup \@sanitize@url \@href}%
\providecommand \@href[1]{\@@startlink{#1}\@@href}%
\providecommand \@@href[1]{\endgroup#1\@@endlink}%
\providecommand \@sanitize@url [0]{\catcode `\\12\catcode `\$12\catcode
  `\&12\catcode `\#12\catcode `\^12\catcode `\_12\catcode `\%12\relax}%
\providecommand \@@startlink[1]{}%
\providecommand \@@endlink[0]{}%
\providecommand \url  [0]{\begingroup\@sanitize@url \@url }%
\providecommand \@url [1]{\endgroup\@href {#1}{\urlprefix }}%
\providecommand \urlprefix  [0]{URL }%
\providecommand \Eprint [0]{\href }%
\providecommand \doibase [0]{http://dx.doi.org/}%
\providecommand \selectlanguage [0]{\@gobble}%
\providecommand \bibinfo  [0]{\@secondoftwo}%
\providecommand \bibfield  [0]{\@secondoftwo}%
\providecommand \translation [1]{[#1]}%
\providecommand \BibitemOpen [0]{}%
\providecommand \bibitemStop [0]{}%
\providecommand \bibitemNoStop [0]{.\EOS\space}%
\providecommand \EOS [0]{\spacefactor3000\relax}%
\providecommand \BibitemShut  [1]{\csname bibitem#1\endcsname}%
\let\auto@bib@innerbib\@empty
\bibitem [{\citenamefont {Smelyanskiy}\ \emph {et~al.}(2020)\citenamefont
  {Smelyanskiy}, \citenamefont {Kechedzhi}, \citenamefont {Boixo},
  \citenamefont {Isakov}, \citenamefont {Neven},\ and\ \citenamefont
  {Altshuler}}]{Smelyanskiy2020}%
  \BibitemOpen
  \bibfield  {author} {\bibinfo {author} {\bibfnamefont {V.~N.}\ \bibnamefont
  {Smelyanskiy}}, \bibinfo {author} {\bibfnamefont {K.}~\bibnamefont
  {Kechedzhi}}, \bibinfo {author} {\bibfnamefont {S.}~\bibnamefont {Boixo}},
  \bibinfo {author} {\bibfnamefont {S.~V.}\ \bibnamefont {Isakov}}, \bibinfo
  {author} {\bibfnamefont {H.}~\bibnamefont {Neven}}, \ and\ \bibinfo {author}
  {\bibfnamefont {B.}~\bibnamefont {Altshuler}},\ }\href {\doibase
  10.1103/PhysRevX.10.011017} {\bibfield  {journal} {\bibinfo  {journal} {Phys.
  Rev. X}\ }\textbf {\bibinfo {volume} {10}},\ \bibinfo {pages} {011017}
  (\bibinfo {year} {2020})}\BibitemShut {NoStop}%
\bibitem [{\citenamefont {Baldwin}\ and\ \citenamefont
  {Laumann}(2018)}]{Baldwin2018}%
  \BibitemOpen
  \bibfield  {author} {\bibinfo {author} {\bibfnamefont {C.~L.}\ \bibnamefont
  {Baldwin}}\ and\ \bibinfo {author} {\bibfnamefont {C.~R.}\ \bibnamefont
  {Laumann}},\ }\href {\doibase 10.1103/PhysRevB.97.224201} {\bibfield
  {journal} {\bibinfo  {journal} {Phys. Rev. B}\ }\textbf {\bibinfo {volume}
  {97}},\ \bibinfo {pages} {224201} (\bibinfo {year} {2018})}\BibitemShut
  {NoStop}%
\bibitem [{\citenamefont {Derrida}(1981)}]{Derrida1981}%
  \BibitemOpen
  \bibfield  {author} {\bibinfo {author} {\bibfnamefont {B.}~\bibnamefont
  {Derrida}},\ }\href {\doibase 10.1103/PhysRevB.24.2613} {\bibfield  {journal}
  {\bibinfo  {journal} {Phys. Rev. B}\ }\textbf {\bibinfo {volume} {24}},\
  \bibinfo {pages} {2613} (\bibinfo {year} {1981})}\BibitemShut {NoStop}%
\bibitem [{\citenamefont {Sherrington}\ and\ \citenamefont
  {Kirkpatrick}(1975)}]{Sherrington1975}%
  \BibitemOpen
  \bibfield  {author} {\bibinfo {author} {\bibfnamefont {D.}~\bibnamefont
  {Sherrington}}\ and\ \bibinfo {author} {\bibfnamefont {S.}~\bibnamefont
  {Kirkpatrick}},\ }\href {\doibase 10.1103/PhysRevLett.35.1792} {\bibfield
  {journal} {\bibinfo  {journal} {Phys. Rev. Lett.}\ }\textbf {\bibinfo
  {volume} {35}},\ \bibinfo {pages} {1792} (\bibinfo {year}
  {1975})}\BibitemShut {NoStop}%
\bibitem [{\citenamefont {Mezard}\ and\ \citenamefont
  {Montanari}(2009)}]{Mezard2009}%
  \BibitemOpen
  \bibfield  {author} {\bibinfo {author} {\bibfnamefont {M.}~\bibnamefont
  {Mezard}}\ and\ \bibinfo {author} {\bibfnamefont {A.}~\bibnamefont
  {Montanari}},\ }\href@noop {} {\emph {\bibinfo {title} {Information, Physics,
  and Computation}}}\ (\bibinfo  {publisher} {Oxford University Press, Inc.},\
  \bibinfo {address} {USA},\ \bibinfo {year} {2009})\BibitemShut {NoStop}%
\bibitem [{\citenamefont {Bapst}\ \emph {et~al.}(2013)\citenamefont {Bapst},
  \citenamefont {Foini}, \citenamefont {Krzakala}, \citenamefont {Semerjian},\
  and\ \citenamefont {Zamponi}}]{Bapst2013}%
  \BibitemOpen
  \bibfield  {author} {\bibinfo {author} {\bibfnamefont {V.}~\bibnamefont
  {Bapst}}, \bibinfo {author} {\bibfnamefont {L.}~\bibnamefont {Foini}},
  \bibinfo {author} {\bibfnamefont {F.}~\bibnamefont {Krzakala}}, \bibinfo
  {author} {\bibfnamefont {G.}~\bibnamefont {Semerjian}}, \ and\ \bibinfo
  {author} {\bibfnamefont {F.}~\bibnamefont {Zamponi}},\ }\href {\doibase
  https://doi.org/10.1016/j.physrep.2012.10.002} {\bibfield  {journal}
  {\bibinfo  {journal} {Physics Reports}\ }\textbf {\bibinfo {volume} {523}},\
  \bibinfo {pages} {127 } (\bibinfo {year} {2013})},\ \bibinfo {note} {the
  Quantum Adiabatic Algorithm Applied to Random Optimization Problems: The
  Quantum Spin Glass Perspective}\BibitemShut {NoStop}%
\bibitem [{\citenamefont {Bouchaud}(1992)}]{Bouchaud1992}%
  \BibitemOpen
  \bibfield  {author} {\bibinfo {author} {\bibfnamefont {J.-P.}\ \bibnamefont
  {Bouchaud}},\ }\href {\doibase 10.1051/jp1:1992238} {\bibfield  {journal}
  {\bibinfo  {journal} {J. Phys. I France}\ }\textbf {\bibinfo {volume} {2}},\
  \bibinfo {pages} {1705} (\bibinfo {year} {1992})}\BibitemShut {NoStop}%
\bibitem [{\citenamefont {Gayrard}(2016)}]{Gayrard2016}%
  \BibitemOpen
  \bibfield  {author} {\bibinfo {author} {\bibfnamefont {V.}~\bibnamefont
  {Gayrard}},\ }\href@noop {} {\enquote {\bibinfo {title} {Aging in
  {M}etropolis dynamics of the {REM}: a proof},}\ } (\bibinfo {year} {2016}),\
  \Eprint {http://arxiv.org/abs/1602.06081} {arXiv:1602.06081 [math.PR]}
  \BibitemShut {NoStop}%
\bibitem [{\citenamefont {Baity-Jesi}\ \emph {et~al.}(2018)\citenamefont
  {Baity-Jesi}, \citenamefont {Biroli},\ and\ \citenamefont
  {Cammarota}}]{Baity-Jesi2018}%
  \BibitemOpen
  \bibfield  {author} {\bibinfo {author} {\bibfnamefont {M.}~\bibnamefont
  {Baity-Jesi}}, \bibinfo {author} {\bibfnamefont {G.}~\bibnamefont {Biroli}},
  \ and\ \bibinfo {author} {\bibfnamefont {C.}~\bibnamefont {Cammarota}},\
  }\href {\doibase 10.1088/1742-5468/aa9f43} {\bibfield  {journal} {\bibinfo
  {journal} {Journal of Statistical Mechanics: Theory and Experiment}\ }\textbf
  {\bibinfo {volume} {2018}},\ \bibinfo {pages} {013301} (\bibinfo {year}
  {2018})}\BibitemShut {NoStop}%
\bibitem [{\citenamefont {Baldwin}\ \emph {et~al.}(2016)\citenamefont
  {Baldwin}, \citenamefont {Laumann}, \citenamefont {Pal},\ and\ \citenamefont
  {Scardicchio}}]{Baldwin2016}%
  \BibitemOpen
  \bibfield  {author} {\bibinfo {author} {\bibfnamefont {C.~L.}\ \bibnamefont
  {Baldwin}}, \bibinfo {author} {\bibfnamefont {C.~R.}\ \bibnamefont
  {Laumann}}, \bibinfo {author} {\bibfnamefont {A.}~\bibnamefont {Pal}}, \ and\
  \bibinfo {author} {\bibfnamefont {A.}~\bibnamefont {Scardicchio}},\ }\href
  {\doibase 10.1103/PhysRevB.93.024202} {\bibfield  {journal} {\bibinfo
  {journal} {Phys. Rev. B}\ }\textbf {\bibinfo {volume} {93}},\ \bibinfo
  {pages} {024202} (\bibinfo {year} {2016})}\BibitemShut {NoStop}%
\bibitem [{\citenamefont {Laumann}\ \emph {et~al.}(2014)\citenamefont
  {Laumann}, \citenamefont {Pal},\ and\ \citenamefont
  {Scardicchio}}]{PhysRevLett.113.200405}%
  \BibitemOpen
  \bibfield  {author} {\bibinfo {author} {\bibfnamefont {C.~R.}\ \bibnamefont
  {Laumann}}, \bibinfo {author} {\bibfnamefont {A.}~\bibnamefont {Pal}}, \ and\
  \bibinfo {author} {\bibfnamefont {A.}~\bibnamefont {Scardicchio}},\ }\href
  {\doibase 10.1103/PhysRevLett.113.200405} {\bibfield  {journal} {\bibinfo
  {journal} {Phys. Rev. Lett.}\ }\textbf {\bibinfo {volume} {113}},\ \bibinfo
  {pages} {200405} (\bibinfo {year} {2014})}\BibitemShut {NoStop}%
\bibitem [{\citenamefont {Faoro}\ \emph {et~al.}(2019)\citenamefont {Faoro},
  \citenamefont {Feigel’man},\ and\ \citenamefont {Ioffe}}]{Faoro2019}%
  \BibitemOpen
  \bibfield  {author} {\bibinfo {author} {\bibfnamefont {L.}~\bibnamefont
  {Faoro}}, \bibinfo {author} {\bibfnamefont {M.~V.}\ \bibnamefont
  {Feigel’man}}, \ and\ \bibinfo {author} {\bibfnamefont {L.}~\bibnamefont
  {Ioffe}},\ }\href {\doibase 10.1016/j.aop.2019.167916} {\bibfield  {journal}
  {\bibinfo  {journal} {Annals of Physics}\ }\textbf {\bibinfo {volume}
  {409}},\ \bibinfo {pages} {167916} (\bibinfo {year} {2019})}\BibitemShut
  {NoStop}%
\bibitem [{\citenamefont {Smelyanskiy}\ \emph {et~al.}(2019)\citenamefont
  {Smelyanskiy}, \citenamefont {Kechedzhi}, \citenamefont {Boixo},
  \citenamefont {Neven},\ and\ \citenamefont {Altshuler}}]{Smelyanskiy2019}%
  \BibitemOpen
  \bibfield  {author} {\bibinfo {author} {\bibfnamefont {V.~N.}\ \bibnamefont
  {Smelyanskiy}}, \bibinfo {author} {\bibfnamefont {K.}~\bibnamefont
  {Kechedzhi}}, \bibinfo {author} {\bibfnamefont {S.}~\bibnamefont {Boixo}},
  \bibinfo {author} {\bibfnamefont {H.}~\bibnamefont {Neven}}, \ and\ \bibinfo
  {author} {\bibfnamefont {B.}~\bibnamefont {Altshuler}},\ }\href@noop {}
  {\enquote {\bibinfo {title} {Intermittency of dynamical phases in a quantum
  spin glass},}\ } (\bibinfo {year} {2019}),\ \Eprint
  {http://arxiv.org/abs/1907.01609} {arXiv:1907.01609 [cond-mat.dis-nn]}
  \BibitemShut {NoStop}%
\bibitem [{\citenamefont {Kravtsov}\ \emph {et~al.}(2015)\citenamefont
  {Kravtsov}, \citenamefont {Khaymovich}, \citenamefont {Cuevas},\ and\
  \citenamefont {Amini}}]{Kravtsov2015}%
  \BibitemOpen
  \bibfield  {author} {\bibinfo {author} {\bibfnamefont {V.~E.}\ \bibnamefont
  {Kravtsov}}, \bibinfo {author} {\bibfnamefont {I.~M.}\ \bibnamefont
  {Khaymovich}}, \bibinfo {author} {\bibfnamefont {E.}~\bibnamefont {Cuevas}},
  \ and\ \bibinfo {author} {\bibfnamefont {M.}~\bibnamefont {Amini}},\ }\href
  {\doibase 10.1088/1367-2630/17/12/122002} {\bibfield  {journal} {\bibinfo
  {journal} {New Journal of Physics}\ }\textbf {\bibinfo {volume} {17}},\
  \bibinfo {pages} {122002} (\bibinfo {year} {2015})}\BibitemShut {NoStop}%
\bibitem [{\citenamefont {Facoetti}\ \emph {et~al.}(2016)\citenamefont
  {Facoetti}, \citenamefont {Vivo},\ and\ \citenamefont
  {Biroli}}]{Facoetti2016}%
  \BibitemOpen
  \bibfield  {author} {\bibinfo {author} {\bibfnamefont {D.}~\bibnamefont
  {Facoetti}}, \bibinfo {author} {\bibfnamefont {P.}~\bibnamefont {Vivo}}, \
  and\ \bibinfo {author} {\bibfnamefont {G.}~\bibnamefont {Biroli}},\ }\href
  {\doibase 10.1209/0295-5075/115/47003} {\bibfield  {journal} {\bibinfo
  {journal} {EPL (Europhysics Letters)}\ }\textbf {\bibinfo {volume} {115}},\
  \bibinfo {pages} {47003} (\bibinfo {year} {2016})}\BibitemShut {NoStop}%
\bibitem [{\citenamefont {Altshuler}\ \emph {et~al.}(2016)\citenamefont
  {Altshuler}, \citenamefont {Cuevas}, \citenamefont {Ioffe},\ and\
  \citenamefont {Kravtsov}}]{Altshuler2016nonergodic}%
  \BibitemOpen
  \bibfield  {author} {\bibinfo {author} {\bibfnamefont {B.~L.}\ \bibnamefont
  {Altshuler}}, \bibinfo {author} {\bibfnamefont {E.}~\bibnamefont {Cuevas}},
  \bibinfo {author} {\bibfnamefont {L.~B.}\ \bibnamefont {Ioffe}}, \ and\
  \bibinfo {author} {\bibfnamefont {V.~E.}\ \bibnamefont {Kravtsov}},\ }\href
  {\doibase 10.1103/PhysRevLett.117.156601} {\bibfield  {journal} {\bibinfo
  {journal} {Phys. Rev. Lett.}\ }\textbf {\bibinfo {volume} {117}},\ \bibinfo
  {pages} {156601} (\bibinfo {year} {2016})}\BibitemShut {NoStop}%
\bibitem [{\citenamefont {Pino}\ \emph {et~al.}(2017)\citenamefont {Pino},
  \citenamefont {Kravtsov}, \citenamefont {Altshuler},\ and\ \citenamefont
  {Ioffe}}]{Pino2017}%
  \BibitemOpen
  \bibfield  {author} {\bibinfo {author} {\bibfnamefont {M.}~\bibnamefont
  {Pino}}, \bibinfo {author} {\bibfnamefont {V.~E.}\ \bibnamefont {Kravtsov}},
  \bibinfo {author} {\bibfnamefont {B.~L.}\ \bibnamefont {Altshuler}}, \ and\
  \bibinfo {author} {\bibfnamefont {L.~B.}\ \bibnamefont {Ioffe}},\ }\href
  {\doibase 10.1103/physrevb.96.214205} {\bibfield  {journal} {\bibinfo
  {journal} {Physical Review B}\ }\textbf {\bibinfo {volume} {96}} (\bibinfo
  {year} {2017}),\ 10.1103/physrevb.96.214205}\BibitemShut {NoStop}%
\bibitem [{\citenamefont {Kravtsov}\ \emph {et~al.}(2018)\citenamefont
  {Kravtsov}, \citenamefont {Altshuler},\ and\ \citenamefont
  {Ioffe}}]{Kravtsov2018}%
  \BibitemOpen
  \bibfield  {author} {\bibinfo {author} {\bibfnamefont {V.}~\bibnamefont
  {Kravtsov}}, \bibinfo {author} {\bibfnamefont {B.}~\bibnamefont {Altshuler}},
  \ and\ \bibinfo {author} {\bibfnamefont {L.}~\bibnamefont {Ioffe}},\ }\href
  {\doibase 10.1016/j.aop.2017.12.009} {\bibfield  {journal} {\bibinfo
  {journal} {Annals of Physics}\ }\textbf {\bibinfo {volume} {389}},\ \bibinfo
  {pages} {148–} (\bibinfo {year} {2018})}\BibitemShut {NoStop}%
\bibitem [{Note1()}]{Note1}%
  \BibitemOpen
  \bibinfo {note} {Note that in the multifractal literature (see {\protect \it
  e.g.\ }Ref.~\cite {RevModPhys.80.1355}), this scaling dimension $D_{\protect
  \mathrm {st}}$ coincides with the fractal dimension $D_1$, also known as the
  ``information dimension''. However, we will make no use of this connection in
  the present work.}\BibitemShut {Stop}%
\bibitem [{\citenamefont {Luca}\ \emph {et~al.}(2013)\citenamefont {Luca},
  \citenamefont {Scardicchio}, \citenamefont {Kravtsov},\ and\ \citenamefont
  {Altshuler}}]{1401.0019}%
  \BibitemOpen
  \bibfield  {author} {\bibinfo {author} {\bibfnamefont {A.~D.}\ \bibnamefont
  {Luca}}, \bibinfo {author} {\bibfnamefont {A.}~\bibnamefont {Scardicchio}},
  \bibinfo {author} {\bibfnamefont {V.~E.}\ \bibnamefont {Kravtsov}}, \ and\
  \bibinfo {author} {\bibfnamefont {B.~L.}\ \bibnamefont {Altshuler}},\
  }\href@noop {} {\enquote {\bibinfo {title} {Support set of random
  wave-functions on the bethe lattice},}\ } (\bibinfo {year} {2013}),\ \Eprint
  {http://arxiv.org/abs/arXiv:1401.0019} {arXiv:1401.0019} \BibitemShut
  {NoStop}%
\bibitem [{\citenamefont {De~Luca}\ \emph {et~al.}(2014)\citenamefont
  {De~Luca}, \citenamefont {Altshuler}, \citenamefont {Kravtsov},\ and\
  \citenamefont {Scardicchio}}]{PhysRevLett.113.046806}%
  \BibitemOpen
  \bibfield  {author} {\bibinfo {author} {\bibfnamefont {A.}~\bibnamefont
  {De~Luca}}, \bibinfo {author} {\bibfnamefont {B.~L.}\ \bibnamefont
  {Altshuler}}, \bibinfo {author} {\bibfnamefont {V.~E.}\ \bibnamefont
  {Kravtsov}}, \ and\ \bibinfo {author} {\bibfnamefont {A.}~\bibnamefont
  {Scardicchio}},\ }\href {\doibase 10.1103/PhysRevLett.113.046806} {\bibfield
  {journal} {\bibinfo  {journal} {Phys. Rev. Lett.}\ }\textbf {\bibinfo
  {volume} {113}},\ \bibinfo {pages} {046806} (\bibinfo {year}
  {2014})}\BibitemShut {NoStop}%
\bibitem [{\citenamefont {Anderson}(1958)}]{Anderson1958}%
  \BibitemOpen
  \bibfield  {author} {\bibinfo {author} {\bibfnamefont {P.~W.}\ \bibnamefont
  {Anderson}},\ }\href {\doibase 10.1103/PhysRev.109.1492} {\bibfield
  {journal} {\bibinfo  {journal} {Phys. Rev.}\ }\textbf {\bibinfo {volume}
  {109}},\ \bibinfo {pages} {1492} (\bibinfo {year} {1958})}\BibitemShut
  {NoStop}%
\bibitem [{Note2()}]{Note2}%
  \BibitemOpen
  \bibinfo {note} {We actually use a linear-in-\( \protect \qopname \relax
  o{log}(t) \) function as our sampled times are logarithmically
  spaced.}\BibitemShut {Stop}%
\bibitem [{\citenamefont {Aizenman}\ and\ \citenamefont
  {Warzel}(2015)}]{aizenman2015random}%
  \BibitemOpen
  \bibfield  {author} {\bibinfo {author} {\bibfnamefont {M.}~\bibnamefont
  {Aizenman}}\ and\ \bibinfo {author} {\bibfnamefont {S.}~\bibnamefont
  {Warzel}},\ }\href {https://books.google.com/books?id=gM9YCwAAQBAJ} {\emph
  {\bibinfo {title} {Random Operators}}},\ Graduate Studies in Mathematics\
  (\bibinfo  {publisher} {American Mathematical Society},\ \bibinfo {year}
  {2015})\BibitemShut {NoStop}%
\bibitem [{Note3()}]{Note3}%
  \BibitemOpen
  \bibinfo {note} {See {\protect \it e.g.\ }Ref \cite {falconer2013fractal},
  chapter 8 for details.}\BibitemShut {Stop}%
\bibitem [{\citenamefont {Zhu}\ \emph {et~al.}(2019)\citenamefont {Zhu},
  \citenamefont {Ochoa},\ and\ \citenamefont
  {Katzgraber}}]{PhysRevE.99.063314}%
  \BibitemOpen
  \bibfield  {author} {\bibinfo {author} {\bibfnamefont {Z.}~\bibnamefont
  {Zhu}}, \bibinfo {author} {\bibfnamefont {A.~J.}\ \bibnamefont {Ochoa}}, \
  and\ \bibinfo {author} {\bibfnamefont {H.~G.}\ \bibnamefont {Katzgraber}},\
  }\href {\doibase 10.1103/PhysRevE.99.063314} {\bibfield  {journal} {\bibinfo
  {journal} {Phys. Rev. E}\ }\textbf {\bibinfo {volume} {99}},\ \bibinfo
  {pages} {063314} (\bibinfo {year} {2019})}\BibitemShut {NoStop}%
\bibitem [{\citenamefont {Thomas}\ and\ \citenamefont
  {Middleton}(2009)}]{PhysRevE.80.046708}%
  \BibitemOpen
  \bibfield  {author} {\bibinfo {author} {\bibfnamefont {C.~K.}\ \bibnamefont
  {Thomas}}\ and\ \bibinfo {author} {\bibfnamefont {A.~A.}\ \bibnamefont
  {Middleton}},\ }\href {\doibase 10.1103/PhysRevE.80.046708} {\bibfield
  {journal} {\bibinfo  {journal} {Phys. Rev. E}\ }\textbf {\bibinfo {volume}
  {80}},\ \bibinfo {pages} {046708} (\bibinfo {year} {2009})}\BibitemShut
  {NoStop}%
\bibitem [{Note4()}]{Note4}%
  \BibitemOpen
  \bibinfo {note} {We remark that even if we leave $\nu $ as a free parameter
  in the fit, we obtain a value of approximately $1/\nu = 2.53$.}\BibitemShut
  {Stop}%
\bibitem [{\citenamefont {Mossi}\ \emph {et~al.}(2017)\citenamefont {Mossi},
  \citenamefont {Parolini}, \citenamefont {Pilati},\ and\ \citenamefont
  {Scardicchio}}]{Mossi_2017}%
  \BibitemOpen
  \bibfield  {author} {\bibinfo {author} {\bibfnamefont {G.}~\bibnamefont
  {Mossi}}, \bibinfo {author} {\bibfnamefont {T.}~\bibnamefont {Parolini}},
  \bibinfo {author} {\bibfnamefont {S.}~\bibnamefont {Pilati}}, \ and\ \bibinfo
  {author} {\bibfnamefont {A.}~\bibnamefont {Scardicchio}},\ }\href {\doibase
  10.1088/1742-5468/aa5286} {\bibfield  {journal} {\bibinfo  {journal} {Journal
  of Statistical Mechanics: Theory and Experiment}\ }\textbf {\bibinfo {volume}
  {2017}},\ \bibinfo {pages} {013102} (\bibinfo {year} {2017})}\BibitemShut
  {NoStop}%
\bibitem [{\citenamefont {Pietracaprina}\ \emph {et~al.}(2018)\citenamefont
  {Pietracaprina}, \citenamefont {Macé}, \citenamefont {Luitz},\ and\
  \citenamefont {Alet}}]{10.21468/SciPostPhys.5.5.045}%
  \BibitemOpen
  \bibfield  {author} {\bibinfo {author} {\bibfnamefont {F.}~\bibnamefont
  {Pietracaprina}}, \bibinfo {author} {\bibfnamefont {N.}~\bibnamefont
  {Macé}}, \bibinfo {author} {\bibfnamefont {D.~J.}\ \bibnamefont {Luitz}}, \
  and\ \bibinfo {author} {\bibfnamefont {F.}~\bibnamefont {Alet}},\ }\href
  {\doibase 10.21468/SciPostPhys.5.5.045} {\bibfield  {journal} {\bibinfo
  {journal} {SciPost Phys.}\ }\textbf {\bibinfo {volume} {5}},\ \bibinfo
  {pages} {45} (\bibinfo {year} {2018})}\BibitemShut {NoStop}%
\bibitem [{\citenamefont {Kechedzhi}\ \emph {et~al.}(2018)\citenamefont
  {Kechedzhi}, \citenamefont {Smelyanskiy}, \citenamefont {McClean},
  \citenamefont {Denchev}, \citenamefont {Mohseni}, \citenamefont {Isakov},
  \citenamefont {Boixo}, \citenamefont {Altshuler},\ and\ \citenamefont
  {Neven}}]{1807.04792}%
  \BibitemOpen
  \bibfield  {author} {\bibinfo {author} {\bibfnamefont {K.}~\bibnamefont
  {Kechedzhi}}, \bibinfo {author} {\bibfnamefont {V.}~\bibnamefont
  {Smelyanskiy}}, \bibinfo {author} {\bibfnamefont {J.~R.}\ \bibnamefont
  {McClean}}, \bibinfo {author} {\bibfnamefont {V.~S.}\ \bibnamefont
  {Denchev}}, \bibinfo {author} {\bibfnamefont {M.}~\bibnamefont {Mohseni}},
  \bibinfo {author} {\bibfnamefont {S.}~\bibnamefont {Isakov}}, \bibinfo
  {author} {\bibfnamefont {S.}~\bibnamefont {Boixo}}, \bibinfo {author}
  {\bibfnamefont {B.}~\bibnamefont {Altshuler}}, \ and\ \bibinfo {author}
  {\bibfnamefont {H.}~\bibnamefont {Neven}},\ }\href@noop {} {\enquote
  {\bibinfo {title} {Efficient population transfer via non-ergodic extended
  states in quantum spin glass},}\ } (\bibinfo {year} {2018}),\ \Eprint
  {http://arxiv.org/abs/arXiv:1807.04792} {arXiv:1807.04792} \BibitemShut
  {NoStop}%
\bibitem [{\citenamefont {Evers}\ and\ \citenamefont
  {Mirlin}(2008)}]{RevModPhys.80.1355}%
  \BibitemOpen
  \bibfield  {author} {\bibinfo {author} {\bibfnamefont {F.}~\bibnamefont
  {Evers}}\ and\ \bibinfo {author} {\bibfnamefont {A.~D.}\ \bibnamefont
  {Mirlin}},\ }\href {\doibase 10.1103/RevModPhys.80.1355} {\bibfield
  {journal} {\bibinfo  {journal} {Rev. Mod. Phys.}\ }\textbf {\bibinfo {volume}
  {80}},\ \bibinfo {pages} {1355} (\bibinfo {year} {2008})}\BibitemShut
  {NoStop}%
\bibitem [{\citenamefont {Falconer}(2013)}]{falconer2013fractal}%
  \BibitemOpen
  \bibfield  {author} {\bibinfo {author} {\bibfnamefont {K.}~\bibnamefont
  {Falconer}},\ }\href {https://books.google.com/books?id=XJN7AgAAQBAJ} {\emph
  {\bibinfo {title} {Fractal Geometry: Mathematical Foundations and
  Applications}}}\ (\bibinfo  {publisher} {Wiley},\ \bibinfo {year}
  {2013})\BibitemShut {NoStop}%
\bibitem [{\citenamefont {Butler}\ \emph {et~al.}(2019)\citenamefont {Butler},
  \citenamefont {Coper}, \citenamefont {Li}, \citenamefont {Lorenzen},\ and\
  \citenamefont {Schopick}}]{Butler2019}%
  \BibitemOpen
  \bibfield  {author} {\bibinfo {author} {\bibfnamefont {S.}~\bibnamefont
  {Butler}}, \bibinfo {author} {\bibfnamefont {E.}~\bibnamefont {Coper}},
  \bibinfo {author} {\bibfnamefont {A.}~\bibnamefont {Li}}, \bibinfo {author}
  {\bibfnamefont {K.}~\bibnamefont {Lorenzen}}, \ and\ \bibinfo {author}
  {\bibfnamefont {Z.}~\bibnamefont {Schopick}},\ }\href@noop {} {\enquote
  {\bibinfo {title} {Spectral properties of the exponential distance matrix},}\
  } (\bibinfo {year} {2019}),\ \Eprint {http://arxiv.org/abs/1910.06373}
  {arXiv:1910.06373 [math.CO]} \BibitemShut {NoStop}%
\bibitem [{Note5()}]{Note5}%
  \BibitemOpen
  \bibinfo {note} {The spectrum of \( {\left (u_{ij}\right )}_{ij} \) can be
  computed analytically using general properties of ``exponential distance
  matrices'' \cite {Butler2019}, showing that the minimum eigenvalue is \(
  (1-e^{-b/n})^n > 0 \).}\BibitemShut {Stop}%
\end{thebibliography}%


%

\appendix
\section{Uniformity in Hamming space}\label{app:U}

In Sec.~\ref{sec:algorithm}, one of the proposed ways to assess the effectiveness of PT is to understand to what extent sampling of the resonant space is performed ``fairly'', \ie without bitstring bias. This corresponds to asking how uniformly the wavefunction is spreading over the microcanonical shell \( \Omega \) as a consequence of time evolution. Here, ``uniformly'' means that the probability distribution ought to explore the entire space \( \Omega \), and not remain concentrated in some corner---\eg the vicinity of the initial state.

It is well-known that the Shannon entropy \( S[p] \) is maximized, with respect to all pdfs defined on a given domain, precisely when \( p \) is the uniform distribution on that domain, so one may think of using \( S \) as a proxy for well-spreadness. However, \( S \) is in all respects insensitive to the \emph{spatial} structure of \( p \): roughly speaking, it only counts on how many states \( p \) is supported, but without care for their mutual Hamming distance.

What we would like to measure instead, is how efficiently PT populates states which are far apart in Hamming distance. To this end, we introduce the ``repulsive potential''
\begin{equation}
    U_\Omega[\psi] = \frac{\sum_{i,j \in \Omega}|\psi_i|^2|\psi_j|^2 u_{ij}}{W_\Omega[\psi]^2},
\end{equation}
where \( W_\Omega[\psi] = \sum_{i\in\Omega}|\psi_i|^2 \) is the total probability of sampling a bitstring inside \( \Omega \) and \( u_{ij} = u(\dist(i,j)) \) is a symmetric, positive-definite ``two-body term'' which decreases with the Hamming distance \( \dist(i,j) \) between \( i \) and \( j \). We take it to be
\begin{equation}\label{eq:uij}
    u_{ij} = e^{-\frac{br_{ij}}{n}},
\end{equation}
where \( r_{ij} = \dist(i,j) \) and \( b \) is an arbitrary positive parameter which we fix to 1.

From a physical perspective, \( U_\Omega \) is akin to an electrostatic potential for the ``charge distribution'' \( p_i = W_\Omega^{-1}|\psi_i|^2 \) defined on \( \Omega \). The denominator ensures that this distribution is properly normalized, so that \( U_\Omega \) does not depend on the behavior of \( |\psi|^2 \) outside the resonant space in the sense that a uniform spillage \( |\psi_i|^2 \mapsto \lambda |\psi_i|^2, \forall i \in \Omega \) (with arbitrary \( \lambda \)) does not change the value of \( U_\Omega \). 

It is easy to see that \( U_\Omega \) is maximized when all the ``charge'' is clumped together, \( p_i = \delta_{i,z_0} \), which corresponds to the fully localized case, \( \Gamma = 0 \). Then \( U_\Omega \) achieves its maximum, \( U^\mathrm{loc}_\Omega = 1 \). Indeed, trying to spread the probability to another site, \( p_i = (1-\varepsilon)\delta_{i,z_0} + \varepsilon\delta_{i,z_1} \), results in a lower potential \( U_\Omega = 1 - 2\varepsilon(1-\varepsilon)(1-u_{z_0,z_1}) < 1 \).

On the opposite end, when \( p_i \) is uniform on \( \Omega \) we expect the functional to be minimal. Indeed, call \( U_0 \) its value,
\begin{equation}\label{eq:U0}
    U_0 = \frac{1}{|\Omega|^2}\sum_{i,j\in\Omega}u_{ij},
\end{equation}
and consider an arbitrary perturbation of the uniform distribution, \( p_i = p_0 + \delta p_i \), where \( p_0 = 1/|\Omega| \) and the perturbation is \emph{nonspilling},
\begin{equation}\label{eq:nonleaking}
    \delta W_\Omega \equiv \sum_{i\in\Omega} \delta p_i = 0.
\end{equation}
This is generic as nonzero values of \( \delta W_\Omega \) can be decomposed into uniform spillage, which does not affect \( U_\Omega \), composed with a nonspilling perturbation, namely \( \delta p_i = \frac{\delta W_\Omega}{|\Omega|} + \delta p_i^\prime \) where \( \delta p_i^\prime \) is nonspilling.

The functional is now written in the form
\begin{equation}\label{eq:U0perturbed}
    U_\Omega[\psi] = U_0 + \frac{2}{|\Omega|}\sum_{i,j\in\Omega}\delta p_i u_{ij} + \sum_{i,j\in\Omega}\delta p_i \delta p_j u_{ij}.
\end{equation}

We now observe that, \emph{on average}, the \( \Omega \) space represents the Hamming cube uniformly (in distribution), namely, we can operate the substitution
\begin{equation}\label{eq:sum-approx}
    \sum_{i,j\in\Omega} = \frac{|\Omega|^2}{V^2}\sum_{i,j}
\end{equation}
when we are interested in the average properties of \( U_\Omega \) (here and throughout, an unspecified summation domain always refers to the entire Hamming cube \( \{0, \dotsc, V-1\} \)). This is because the REM has totally uncorrelated energy levels, so the two-point indicator factorizes: \( \operatorname{\mathbb{E}}\left[\chi_\Omega(i)\chi_\Omega(j)\right] = \operatorname{\mathbb{E}}\left[\chi_\Omega(i)\right]\operatorname{\mathbb{E}}\left[\chi_\Omega(j)\right] = |\Omega|^2/V^2 \).

Therefore, when we rewrite the sum in the second term at the r.h.s.\ of Eq.~\eqref{eq:U0perturbed} as
\begin{equation}
    \sum_{i,j\in\Omega}\delta p_i u_{ij} = \sum_{i\in\Omega}\delta p_i \sum_{j\in\Omega} u_{ij},
\end{equation}
we recognize that the inner sum \( \sum_{j\in\Omega} u_{ij} = \frac{|\Omega|}{V}\sum_j u_{ij} \) (equality in expectation values) does not actually depend on \( i \), since \( \sum_j u_{ij} = \sum_{r=0}^n {n\choose r} u(r) \).
As a consequence, the whole term vanishes due to Eq.~\eqref{eq:nonleaking}.

This, along with the fact that the third term at the r.h.s.\ of Eq.~\eqref{eq:U0perturbed} is a quadratic form of the positive definite matrix Eq.~\eqref{eq:uij} \footnote{The spectrum of \( {\left(u_{ij}\right)}_{ij} \) can be computed analytically using general properties of ``exponential distance matrices'' \cite{Butler2019}, showing that the minimum eigenvalue is \( (1-e^{-b/n})^n > 0 \).}, implies that \( U_\Omega[\psi] \ge U_0 \), with equality only in the uniform case \( \delta p_i \equiv 0 \).

This indicates that \( U_\Omega \) can be used as a measure of uniformity of the wavefunction intensity distribution. We conclude by estimating the specific value of \( U_0 \) expected of a totally extended system.

From Eqs.~\eqref{eq:U0} and \eqref{eq:sum-approx}, we have
\begin{multline}
    U_0 = \frac{1}{V^2}\sum_{i,j} e^{-\frac{br_{ij}}{n}} = \frac{1}{V}\sum_{r=0}^n e^{-\frac{br}{n}} \\
    = \left(\frac{1+e^{-b/n}}{2}\right)^n \sim e^{-b/2},
\end{multline}
in good accord with both the NEE and ergodic curves in Fig.~\ref{fig:spread_spillage}.

\end{document}